\documentclass[acmsmall,screen]{acmart}

\usepackage{colortbl}
\usepackage{makecell}
\usepackage{arydshln}
\usepackage{hyperref}
\usepackage{color}
\usepackage{bigstrut}
\usepackage[ruled]{algorithm2e}
\usepackage{amsmath,amsfonts}
\usepackage{amsmath}

\usepackage{algorithmic}
\usepackage{balance}
\usepackage{blindtext}
\usepackage{booktabs}
\usepackage[noabbrev]{cleveref}
\usepackage{comment}
\usepackage{fancyvrb}
\usepackage{framed}
\usepackage{graphicx}
\usepackage{ifthen}
\usepackage{listings}
\usepackage{mathrsfs}
\usepackage{mdwmath}
\usepackage{multirow}
\usepackage{pifont}
\usepackage{textcomp}
\usepackage{url}
\usepackage{xspace}
\usepackage[normalem]{ulem}
\usepackage[framemethod=TikZ]{mdframed}
\usepackage{caption}
\usepackage{subcaption}
\usepackage{tcolorbox}
\tcbuselibrary{listings,skins}
\usepackage{mathtools}
\usepackage{blindtext}
\usepackage{lstautogobble}
\usepackage{wrapfig}

\DeclareGraphicsExtensions{.pdf,.jpeg,.png}
\graphicspath{{figures/}}

\newcommand{\rom}[1]{\uppercase\expandafter{\romannumeral #1\relax}}

\newcommand{\eg}{\hbox{\emph{e.g.,}}\xspace}
\newcommand{\ie}{\hbox{\emph{i.e.,}}\xspace}

\newlength\Linewidth
\def\findlength{\setlength\Linewidth\linewidth
\addtolength\Linewidth{-4\fboxrule}
\addtolength\Linewidth{-3\fboxsep}
}
\newenvironment{examplebox}{\par\begingroup
  \setlength{\fboxsep}{3pt}\findlength
  \setbox0=\vbox\bgroup\noindent
  \hsize=0.90\linewidth
  \begin{minipage}{0.90\linewidth}\normalsize}
    {\end{minipage}\egroup
    \textcolor{gray20}{\fboxsep1.2pt\fbox
     {\fboxsep5pt\colorbox{gray05}{\normalcolor\box0}}}
    \endgroup\par\noindent
    \normalcolor\ignorespacesafterend}

\usetikzlibrary{shadows}
\usepackage{graphics}
\newmdenv[
    tikzsetting= {fill=blueish},
    skipabove=0.33em,
    skipbelow=0.33em,
    linewidth=1pt,
    innerleftmargin=4pt,
    innerrightmargin=4pt,
    innertopmargin=2pt,
    innerbottommargin=2pt,
    linecolor=gray95,
    roundcorner=2pt, 
    shadow=true,
    shadowsize=4pt,
    shadowcolor=gray95
]{questionbox}

\newmdenv[
    tikzsetting= {fill=greenish},
    skipabove=0.33em,
    skipbelow=0.33em,
    linewidth=1pt,
    innerleftmargin=4pt,
    innerrightmargin=4pt,
    innertopmargin=2pt,
    innerbottommargin=2pt,
    linecolor=gray95,
    roundcorner=2pt, 
    shadow=true,
    shadowsize=4pt,
    shadowcolor=gray95
]{answerbox}

\newmdenv[
    skipabove=0.33em,
    skipbelow=0.33em,
    innerleftmargin=4pt,
    innerrightmargin=4pt,
    innertopmargin=2pt,
    innerbottommargin=2pt,
]{lessonbox}

\usepackage{tikz}

\definecolor{javared}{rgb}{0.6,0,0} %
\definecolor{javagreen}{rgb}{0.25,0.5,0.35} %
\definecolor{javapurple}{rgb}{0.5,0,0.35} %
\definecolor{javadocblue}{rgb}{0.25,0.35,0.75} %

\lstdefinestyle{basejava}{
  language=java,
  showstringspaces=false,
  basicstyle=\small\ttfamily,
  keywordstyle=\bfseries\color{javapurple},
  commentstyle=\itshape\blue,
  identifierstyle=\blue,
  frame=none,
  backgroundcolor=\color{white},
}

\lstdefinestyle{CustomJava}{
  belowcaptionskip=\baselineskip,
  breaklines=true,
  xleftmargin=\parindent,
  language=java,
  showstringspaces=false,
  basicstyle=\scriptsize\ttfamily,
  keywordstyle=\bfseries\color{javapurple},
  commentstyle=\itshape\blue,
  identifierstyle=\blue,
  belowskip=1pt,
  numbers=left,
  gobble=0
}

\lstdefinestyle{CustomJavaWoNumbers}{
  belowcaptionskip=0.5\baselineskip,
  breaklines=true,
  xleftmargin=\parindent,
  language=java,
  showstringspaces=false,
  basicstyle=\scriptsize\ttfamily,
  keywordstyle=\bfseries\color{javapurple},
  commentstyle=\itshape\blue,
  identifierstyle=\blue,
  belowskip=0.5pt,
  numbers=none,
  gobble=0
}

\newcommand\blue[1]{\textcolor[rgb]{0.00,0.00,1.00}{{#1}}}

\definecolor{blueish}{RGB}{250, 250, 255}
\definecolor{greenish}{RGB}{250, 255, 250}
\definecolor{redish}{RGB}{255, 200, 200}

\definecolor{highlight}{RGB}{175, 255, 100}
\definecolor{gray01}{gray}{.98}
\definecolor{gray05}{gray}{0.95}
\definecolor{gray08}{gray}{0.92}
\definecolor{gray10}{gray}{0.90}
\definecolor{gray12}{gray}{0.88}
\definecolor{gray15}{gray}{0.85}
\definecolor{gray18}{gray}{0.82}
\definecolor{gray20}{gray}{0.80}
\definecolor{gray25}{gray}{0.75}
\definecolor{gray30}{gray}{0.70}
\definecolor{gray35}{gray}{0.65}
\definecolor{gray40}{gray}{0.60}
\definecolor{gray45}{gray}{0.55}
\definecolor{gray50}{gray}{0.50}
\definecolor{gray55}{gray}{0.45}
\definecolor{gray60}{gray}{0.40}
\definecolor{gray65}{gray}{0.35}
\definecolor{gray70}{gray}{0.30}
\definecolor{gray75}{gray}{0.25}
\definecolor{gray80}{gray}{0.20}
\definecolor{gray85}{gray}{0.15}
\definecolor{gray90}{gray}{0.10}
\definecolor{gray95}{gray}{0.05}
\definecolor{rowgray}{RGB}{224, 224, 224}

\newcommand{\tool}{\textsc{Cycle}\xspace}

\newcommand{\improve}[1]{{\textcolor{red}{\scriptsize{$#1$}}}}
\newcommand{\blimprove}[1]{{\textcolor{gray}{\scriptsize{$#1$}}}}

\newcommand{\coloryellow}{\cellcolor[rgb]{1,0.925,0.792}}

\newcommand{\increase}[1]{{\textcolor{javagreen}{\scriptsize{$#1$}}}}
\newcommand{\decrease}[1]{{\textcolor{red}{\scriptsize{$#1$}}}}

\newcommand{\colorgray}{\cellcolor[rgb]{0.906, 0.902, 0.902}}
\newcommand{\grayline}{\rowcolor[gray]{.90}}

\newtcbox{\inlinebox}[1][]{enhanced,
 box align=base,
 nobeforeafter,
 colback=blueish,
 size=small,
 left=0pt,
 right=0pt,
 boxsep=2pt,
 #1}

\renewcommand{\cref}[1]{\Cref{#1}}

\usepackage{tcolorbox}
\newcounter{findingCounter}
\newenvironment{finding}{
\begin{tcolorbox}[colback=blue!5!white,colframe=blue!5!white,arc=0mm,grow to left by=0mm,left=0mm,grow to right by=0mm,left=1.5mm,right=1.5mm,top=1.5mm,bottom=1.5mm]
\textbf{Result-\arabic{findingCounter}\stepcounter{findingCounter}:}
}
{
\end{tcolorbox}
}

\lstdefinelanguage{c-pretty}
{
  language=c,
  numbers=left,
  basicstyle=\footnotesize\ttfamily,
  numberstyle=\footnotesize,
  breaklines=true,
  columns=fullflexible,
  xleftmargin=0pt,
  showstringspaces=false,
  identifierstyle=\color{black},
  keywordstyle=\color{javapurple}\bfseries,
  stringstyle=\color{javared},
  commentstyle=\color{javagreen},
  morecomment=[s][\color{javadocblue}]{/**}{*/},
}
\acmJournal{PACMPL}
\acmVolume{1}
\acmNumber{CONF} %
\acmArticle{1}
\acmYear{2018}
\acmMonth{1}
\acmDOI{10.1145/nnnnnnn.nnnnnnn}
\startPage{1}

\setcopyright{rightsretained}
\acmJournal{PACMPL}
\acmYear{2024} \acmVolume{8} \acmNumber{OOPSLA1} \acmArticle{108} \acmMonth{4}\acmDOI{10.1145/3649825}
\begin{document}

\title{CYCLE: Learning to Self-Refine the Code Generation}

\author{Yangruibo Ding}
\email{}
\affiliation{%
  \institution{Columbia University}
  \city{New York}
  \state{NY}
  \country{USA}
}

\author{Marcus J. Min}
\affiliation{%
  \institution{Columbia University}
  \city{New York}
  \state{NY}
  \country{USA}
}

\author{Gail Kaiser}
\affiliation{%
  \institution{Columbia University}
  \city{New York}
  \state{NY}
  \country{USA}
}

\author{Baishakhi Ray}

\affiliation{%
  \institution{Columbia University}
  \city{New York}
  \state{NY}
  \country{USA}
}

\begin{abstract}
  Pre-trained code language models have achieved promising performance in code generation and improved the programming efficiency of human developers. However, their self-refinement capability is typically overlooked by the existing evaluations of code LMs, which focus only on the accuracy of the one-time prediction. For the cases when code LMs fail to implement the correct program, developers actually find it hard to debug and fix the faulty prediction since it is not written by the developers themselves. Unfortunately, our study reveals that code LMs cannot efficiently self-refine their faulty generations as well. 
  
  In this paper, we propose \tool framework, learning to self-refine the faulty generation according to the available feedback, such as the execution results reported by the test suites. We evaluate \tool on three popular code generation benchmarks, HumanEval, MBPP, and APPS. The results reveal that \tool successfully maintains, sometimes improves, the quality of one-time code generation, while significantly improving the self-refinement capability of code LMs. We implement four variants of \tool with varied numbers of parameters across 350M, 1B, 2B, and 3B, and the experiments show that \tool consistently boosts the code generation performance, by up to 63.5\%, across benchmarks and varied model sizes. We also notice that \tool outperforms code LMs that have 3$\times$ more parameters in self-refinement.

\end{abstract}


\maketitle

\section{Introduction}

Pre-trained code language models (code LMs) have achieved great success in code generation, and many of them have been deployed as a part of the integrated development environment (IDE), such as GitHub Copilot~\citep{github-2021-copilot} and Amazon CodeWhisperer~\citep{amazon-2023-codewhisperer}, to help human developers improve the programming efficiency. Along this direction, researchers started to conduct empirical and human studies to analyze the strengths and weaknesses of these models~\citep{barke2023grounded, guo2023exploring, huang2023large}. For example, \citet{barke2023grounded} propose a grounded theory of code-LM-assisted programming, systematically categorizing the interaction between code LMs and developers into two modes: \emph{acceleration} and \emph{exploration}. They define the \emph{acceleration} mode as the situation when the developers clearly know what are the expected functionalities, code LMs help them speed up the implementation. On the other hand, the \emph{exploration} mode indicates the scenario when developers do not have concrete plans regarding how to proceed, such as facing some bugs reported by the test suites while developers have not figured out how to fix them. While this study identifies the efficiency of code LMs in the acceleration mode, it recognizes their limitations in exploration mode. For example, the code generated by code LMs in the exploration mode is less trusted than in the acceleration mode, and developers tend to validate them frequently with executions. 
Moreover, when errors are revealed through executions, developers struggle to debug model-generated code, as it was not authored by themselves, making it more challenging to understand the error within unfamiliar code.

In this paper, we propose \tool framework, making an attempt to enhance the performance of code LMs in the exploration mode.
The fundamental principle underpinning the creation of \tool is the recognition that expecting code LMs to excel in the exploration mode, where human intentions are often unclear or not explicitly specified, may be overly demanding. However, these models should possess the capability to iteratively improve their code generation based on the feedback they receive from other sources such as execution results reported by test suites. In essence, \tool aims to empower code LMs to adapt and enhance their output in response to the available feedback, thereby bridging the gap between human developers' exploratory programming needs and the capabilities of code LMs.

\paragraph{\textbf{Limitations of code LMs in the exploration mode}} In this work, we focus on the scenario of code generation that given the natural language (NL) description of a problem, typically wrapped in the docstring, the code LM will implement the program accordingly. For the convenience of our future discussion, we first concretize the \emph{acceleration} mode and \emph{exploration} mode in our scenario.

\begin{itemize}
    \item Acceleration Mode: Given the NL description of a problem, code LMs directly predict the code accordingly.
    \item Exploration Mode: If the prediction from the acceleration mode fails the test cases and execution feedback returns, code LMs try to refine the faulty code without further human instructions. 
\end{itemize}

\begin{figure*}
    \centering
    \includegraphics[width=0.98\textwidth]{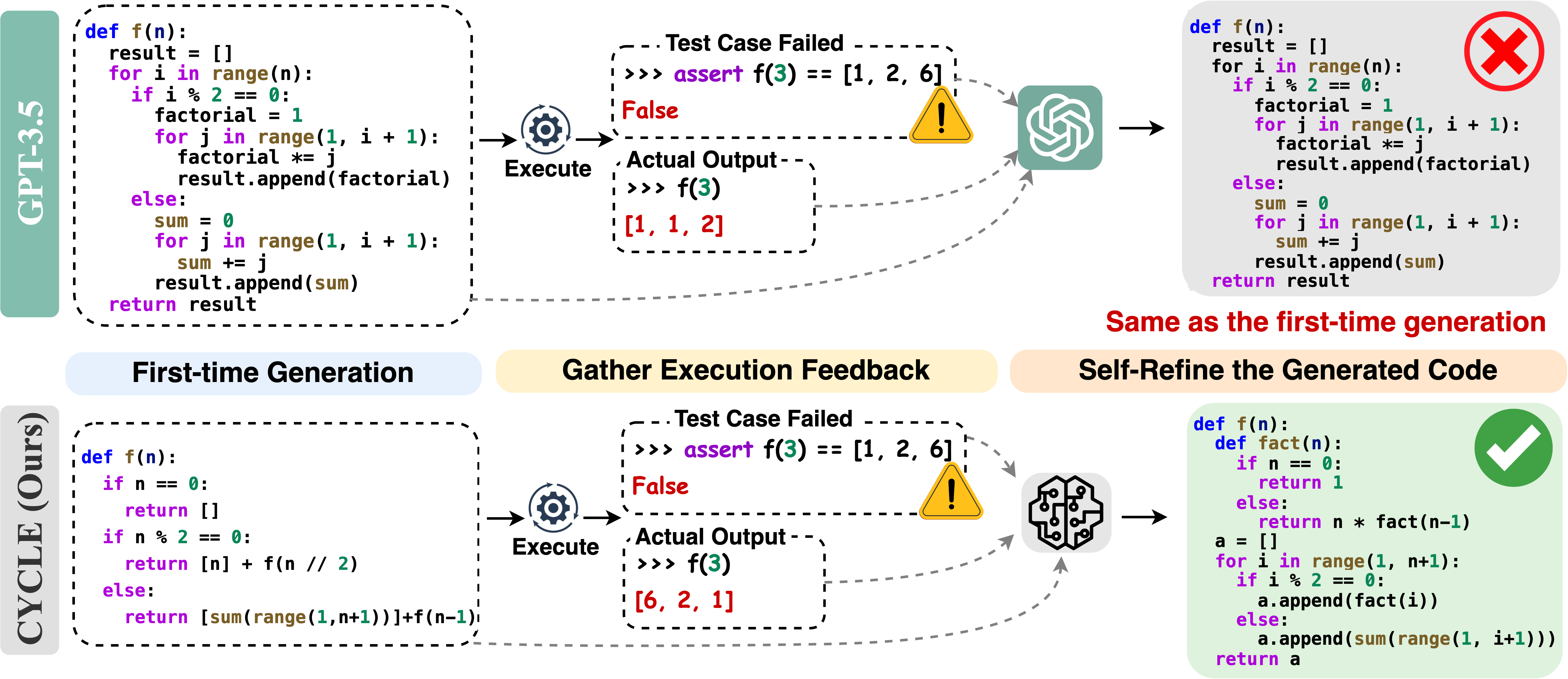}
    \caption{Motivation Example. We prompt GPT-3.5 and \tool to implement a program according to a problem description from the HumanEval programming benchmark (Task No. 106). While both failed to pass the test suite in the acceleration mode, \tool successfully refined its own generation referring to the execution feedback. In contrast, GPT-3.5 could not self-refine effectively.}
    \label{fig:motivation}
    \vspace{2mm}
\end{figure*}

As shown in Figure~\ref{fig:motivation}, we prompt GPT-3.5~\citep{ouyang2022instructgpt} with a problem description from HumanEval programming benchmark~\citep{chen2021codex}. GPT-3.5 could not fulfill the functionality in the acceleration mode, and its generated code failed to pass the test suite of this problem. We could see from GPT-3.5's generation that, though the program is very short, spanning only 14 lines, it is not straightforward for humans to understand. The complexity mainly comes from the nested for-loops, making it even more difficult to manually identify and correct the error. Therefore, motivated by ~\citet{chen2023teaching}, we tried concatenating the faulty generation and the execution feedback reported by the test suite, as additional references, with the problem description and expected the model to refine the generated code by itself in the exploration mode. Unfortunately, GPT-3.5 could not effectively understand the guidance from the execution feedback and simply copy-pasted the faulty code as its new prediction. 

Such weakness of self-refinement in the exploration mode is even more severe in open-source code LMs. We conduct similar experiments with CodeGen (2.7 billion parameters)~\citep{nijkamp2023codegen} and StarCoder (3 billion parameters)~\citep{li2023starcoder} on the whole HumanEval benchmark with 164 programming problems. We observe that existing code LMs perform poorly in the exploration mode, failing to self-refine the faulty generations according to the execution feedback. CodeGen generates an exact copy of the faulty code as its refined prediction in 42.2\% cases while StarCoder copies in 64.8\% cases. Such weak self-refinement capability in the exploration mode is concerning, as it brings further burden to the human developers to fix the bugs brought by the model-generated code.

\paragraph{\textbf{Our Approach}}~~
In this work, we argue that code LMs should be enhanced in exploration Mode with self-refinement capability by leveraging the available feedback from the execution results. In fact, a model trained with such execution feedback has the potential to perform better even in the acceleration mode. %

To this end, we design \tool, a framework that teaches code LMs to self-refine (by continuing training a pre-trained model) by jointly attending to three information sources: 
(i) the high-level problem description in natural language, 
(ii) the incorrect code the model may have generated in the previous attempts, and 
(iii) the execution feedback. 
We have developed an input template that consolidates these three information sources and employs them to train the code LM. While traditional code LMs' training primarily relies on the first information source, the inclusion of prior incorrect code aids the model in achieving a more comprehensive grasp of its own errors. The execution feedback, in turn, guides the code LM in generating programs that align precisely with the problem description.

Nonetheless, when we naively include the previously generated erroneous code in the input, the code LM often resorts to a shortcut, essentially copying from the incorrect input when generating new code. To deter the code LM from adopting such shortcuts, we employ a masking technique, referred to as the Past Generation Mask (PGM). This strategy slightly obfuscates the incorrect past generations, motivating the model to explore a more extensive range of solutions for code refinement. Additionally, to strike a balance between the proficiency of code generation in acceleration and exploration mode, we employ a data mixing strategy that manipulates the ratio of self-refinement features and general code completion features.

To efficiently train a Code LM using the above strategy we have to curate data that simulate the exploration mode development. 
Thus, we further design an automatic training data generation phase as 
existing pre-training code datasets~\citep{xu2022polycoder, kocetkov2022stack, nijkamp2023codegen} are challenging to be customized for self-refinement training. Our data collection phase automatically prompts the pre-trained code LMs to reveal their own strengths and weaknesses in code generation, verified by executing test cases, and constructs data samples to reinforce their strengths while refining their weaknesses.

Finally, we implement \tool to realize an automated self-refinement workflow that imitates the iterative programming practice of human developers. The workflow first prompts code LMs to initialize the implementation according to the high-level problem description and then continuously verifies the correctness of prediction with execution and aggregates comprehensive information for further refinement.

\paragraph{\textbf{Results}}~~We evaluate \tool's code generation capability with three popular programming benchmarks: HumanEval~\citep{chen2021codex}, MBPP-Sanitized~\citep{austin2021synthesis}, and APPS~\citep{hendrycks2021apps}. To illustrate the effectiveness and generalizability of \tool, we train four variants of \tool with varied parameter sizes ranging from 350M to 3B. From the evaluation results, we conclude that \tool is pretty effective at self-refinement, consistently boosting the code generation performance, by up to 63.5\% relative improvement, across four model sizes on all three benchmarks, while maintaining decent one-time generation capacity. With efficient self-refinement learning, \tool-350M outperforms StarCoder-1B across all three programming benchmarks, and \tool-1B matches the performance of StarCoder-3B. With the in-depth analysis, we also empirically reveal that \tool is effective at capturing execution feedback and has great potential to assist human developers with iterative programming.

\paragraph{\textbf{Novelty and Contributions.}} We make the following novel contributions:
\begin{itemize}
\item Our work sheds light on the weaknesses of code LMs in self-refinement, revealing that these models are not effective at understanding the execution feedback and correcting their own mistakes accordingly. 

\item To fulfill the code LMs' potential in self-refinement, we propose \tool, a framework that enhances the code LMs' generation performance by learning to refine their own generated code. We first propose a knowledge-distillation-based data collection approach to automatically construct samples to teach code LMs to self-refine. We then propose a training strategy designed specifically for learning self-refinement. Finally, we implement an iterative self-refinement workflow that automates the process of generating code in exploration mode. 

\item We conduct extensive experiments on three popular code generation benchmarks with four \tool variants across 350M to 3B model parameters, and show that \tool consistently increases the code generation performance by up to 63.5\%. \tool could also match or even outperform baseline code LMs with 3$\times$ parameters.

\item We perform in-depth analysis to discuss \tool's design and performance from multiple perspectives. We conclude with insights and takeaways to motivate further research in improving code LM's self-refinement capability, which hopefully assists human developers with iterative programming and generally increases code LM's performance in exploration mode.

\end{itemize}

We anonymously release our code, data, and model checkpoints. The artifact is available at \url{https://github.com/ARiSE-Lab/CYCLE_OOPSLA_24}.
\section{Overview}

\begin{figure*}
    \centering
    \includegraphics[width=0.98\textwidth]{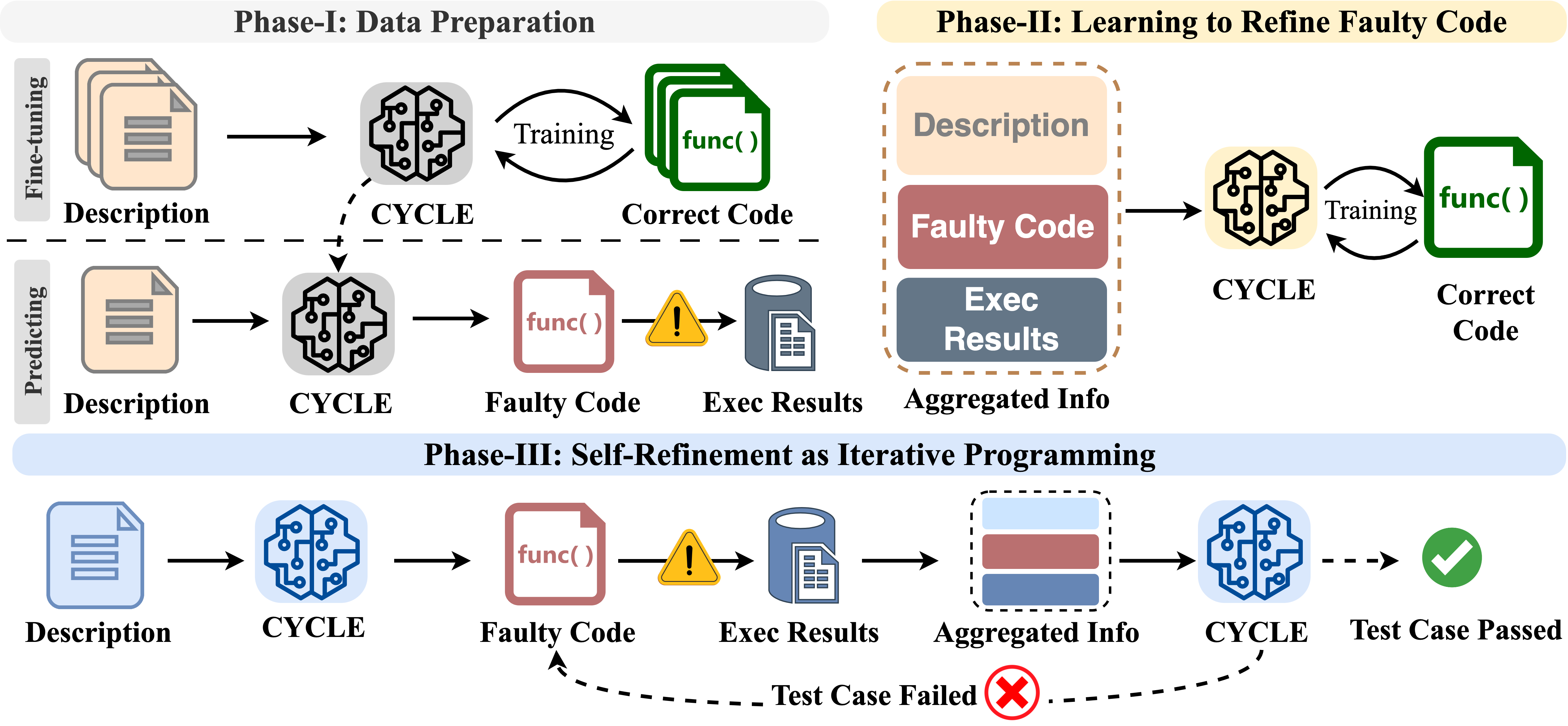}
    \caption{Overview of \tool.}
    \label{fig:overview}
\end{figure*}

In this section, we briefly introduce \tool, explaining the high-level designs and the intuitions behind them. We present the overview of \tool in Figure~\ref{fig:overview}. At a high level, \tool contains three phases. 

\paragraph{Phase-I: Data Preparation for Self-Refinement} The data to train code LMs for refinement capability requires carefully crafted features, such as the developing log with iterative correction of code snippets and their error-exposing feedback loops. These features do not naturally come along with the large-scale pre-training datasets~\cite{xu2022polycoder, kocetkov2022stack, nijkamp2023codegen} that typically prioritize data quantity while not providing interfaces for customization. Therefore, we propose an automated approach to distill such features from the pre-trained code LMs and construct the datasets on top of them. 

The general idea is to prompt the pre-trained code LMs with programming problems, asking them to generate code to fulfill the requested functionalities. These problems should be well-defined and accompanied by test suites and canonical solutions so that the model's generation can be efficiently verified. When the code LM makes mistakes, we gather its faulty generations and the corresponding execution feedback reported by the test suites. Then we construct data samples accordingly, which could be leveraged to teach code LMs to correct their own mistakes by referring to the execution feedback and the canonical solutions accompanied by the programming challenges. 

\noindent\textbf{Fine-tuning Code LMs with Correct Code}.~~Code LMs are pre-trained with up to trillions of code tokens, but the noise inherited in the pre-training data could 
trigger unexpected predictions~\citep{li2022alphacode} or even vulnerable code~\citep{sven-llm}. We should avoid these malicious behaviors when constructing samples for self-refinement. The unexpected predictions might reveal random errors and could not get meaningful feedback after execution. Also, when the vulnerable code is generated, it might bring a security threat to the execution system. Therefore, we propose to first fine-tune the pre-trained code LMs on the correct code to minimize the model's malicious behaviors. Concretely, we collect a decent number of programming challenges that are accompanied by canonical solutions, and we fine-tune the pre-trained code LMs to predict these guaranteed correct programs conditioned on the problem description.

\noindent\textbf{Prompt Code LMs to Distill the Weaknesses}.~~With the fine-tuning on the canonical solutions, we expect the model to make faulty predictions that are not too off from the correct code. We prompt the fine-tuned model with the problem description and then execute its prediction with the accompanied test suites. The execution results with error messages expose the models' strengths and weaknesses in code generation, which are collected and saved as resources to build training samples to learn the refinement.

\paragraph{Phase-II: Learning to Refine the Faulty Generation} With the data prepared by Phase-I, we construct the training samples to learn code refinement based on three types of information. Specifically, we aggregate the problem description, the faulty code generated by the code LM, and the corresponding execution results, using our proposed template~(more details in Section~\ref{subsubsec:align_multiple_resources}), to formulate the model input, so that the model could jointly attend to comprehensive information simultaneously. Then we use the accompanying canonical solutions as the target for the model to predict. Different from the fine-tuning in Phase-I, which only teaches the model to fulfill the functionalities described in natural language, the code refinement training in this phase exposes the model to its own mistakes and the execution feedback, together with the problem description, which forces the model to both reason about the misalignment between its generation in the past and the problem description and learn to understand the implicit guidance in the execution results. We carefully design our training strategy to effectively and efficiently learn the code refinement, and we will introduce more details of our approach in Section~\ref{subsec:self_refine_training}.

\paragraph{Phase-III: Self-Refinement as Iterative Programming} After learning about the code refinement from Phase-II, we deploy the model to automatically generate code according to the problem description and, similar to the iterative programming practice of human developers, iteratively refine the code to fulfill the required functionality. When the problem description is fed into the model, the model will first generate the code at its best, and the generation will automatically be executed with the test suite. If failed test cases are detected, our framework will automatically aggregate the description, the faulty code, and the execution feedback using the template proposed in Phase-II. Finally, the aggregated information will be again fed into the model for self-refinement.

\section{Approach}
\label{sec:approach}
In this section, we explain \tool in detail. We will illustrate the concrete approaches we designed to teach the code language models (code LMs) about self-refinement for code generation.

\subsection{Phase-I: Data Preparation}
\label{subsec:data_preparation}
Training code LMs to possess self-refinement capabilities demands a dataset rich in specific features, such as development logs showcasing the iterative correction of code snippets, as well as feedback loops that highlight the errors. Existing large-scale pre-training datasets~\citep{xu2022polycoder, nijkamp2023codegen, kocetkov2022stack} are primarily designed to amass vast amounts of data but are not inherently equipped to offer these nuanced features. Therefore, an efficient data collection method is required to extract these specialized features. To this end, we propose a knowledge distillation~\citep{west-etal-2022-symbolic} approach that prompts the pre-trained code LMs to showcase their capabilities, revealing their strength as well as exposing their weaknesses. Subsequently, we will construct the datasets to reinforce their strengths while learning to self-refine their weaknesses.

\subsubsection{Fine-tune Code LMs with Semantically Correct Code} 
\label{subsubsec:approach_ft_w_correct_code}
Code language models~\citep{li2023starcoder, chen2021codex, rozière2023codellama, nijkamp2023codegen} are typically pre-trained with a huge amount of source code, learning to predict up to trillions of code tokens. However, the innate noise in the pre-training data could affect its accuracy in code generation. As the large corpora are mostly collected from open-source resources~\citep{kocetkov2022stack, rozière2023codellama, xu2022polycoder}, the quality of training samples varies significantly. Specifically, the high-quality code snippets well align with their accompanying natural-language docstrings or comments, such as those from mature and actively maintained projects or forked commercial software, while a certain amount of the noisy snippets have vulnerable functionalities and semantic misalignment, such as those from developing projects or starter-level programmers. This means that during training, the model is exposed to, and possibly memorizes, both correct (aligned) and incorrect (misaligned) programs. The consequential challenge during the knowledge distillation is that, given a natural language (NL) prompt, the model holds the potential to generate either semantically accurate or erroneous code. This behavior is generally determined by the frequency ratio of correct to incorrect code observed during pre-training. It has been verified by existing works that the noise in the pre-training data could bring unexpected behavior~\citep{li2022alphacode} or security vulnerabilities~\citep{sven-llm}.

To minimize the malicious behaviors of code LMs during the knowledge distillation, we first fine-tune the pre-trained code LMs using only ``guaranteed correct" code. Specifically, we collect canonical solutions from programming challenges that are already verified by the test suite and fine-tune code LMs to predict these solutions token by token. Different from the pre-training of code LMs~\citep{rozière2023codellama, li2023starcoder, nijkamp2023codegen, nijkamp2023codegen2} that predicts every token in the corpora, our fine-tuning is designed to only limited to the code tokens within the canonical solution and the NL description will be regarded only as context. Specifically, the NL description is a sequence of $m$ tokens, $NL = \{nl_0, nl_1, ..., nl_m\}$, and the canonical solution is a sequence of $n$ tokens, $C = \{c_0, c_1, ..., c_n\}$, we apply the standard language modeling loss~\citep{Radford2018gpt1, Radford2019gpt2, brown2020language} to optimize the fine-tuning:

\begin{equation}
\label{eq:mlm_loss}
    \mathcal{L}_{fine-tune} = \sum_{n \in |C|} - log P(c_n ~| ~NL, c_1, c_2, ..., c_{n-1})
\end{equation}

Such a design makes the model's prediction more focused, and the learning process is more narrowed to only optimize the model towards the code generation. We observe that, after this fine-tuning, the model not only generates better code quality but also 
knows when to terminate the generation more accurately than the pre-trained code LMs without such fine-tuning. This is mainly because the pre-training asks code LMs to predict tokens until the maximum context length is reached, while we fine-tune the model to focus on the canonical solutions that are terminated naturally when the functionality is complete. 

\subsubsection{Prompt Code LMs to Distill the Weaknesses}
\label{subsubsec:distill_weakness} After the fine-tuning, the model is primed to generate solutions that, even if faulty, are proximate to canonical solutions. This leaves us great chances to create a dataset with a decent amount of paired samples. Such a pair, including a correct and a wrong solution targeting the same problem, is quite valuable, revealing through error messages the subtle nuances where the model's code generation aligns or diverges from the desired outcome. These errors aren't merely mistakes; they provide insights into the model's comprehension and interpretation, shedding light on its strengths and weaknesses. We hope to teach code LMs to capture and learn to transform the wrong code to its correct counterpart by fixing the subtle error according to the execution feedback. Such a transformation well imitates the process of self-refinement. 

To construct such samples, we prompt code LMs with problem descriptions and verify their predictions using the accompanying test suites. For those problems that the model fails to predict correctly, we gather its faulty generation and the errors thrown by the test suite execution. We will pair this fault with the ground-truth implementation to construct training samples for the next phase of learning. For the correctly predicted problem, we will also gather its generation to replace the canonical solution of this problem. For the next phase of learning, we will train the model to predict its own generated correct code to maintain and reinforce such knowledge.

\subsection{Phase-II: Learning to Refine the Faulty Generation}
\label{subsec:self_refine_training}
With the constructed samples in the previous phase, we introduce the training process of \tool, which is designed for learning to refine the faulty code.

\tool will be firstly initialized with the fine-tuned checkpoints that we introduced in Section~\ref{subsubsec:approach_ft_w_correct_code}, and then continue training with the samples described in Section~\ref{subsubsec:distill_weakness}. As these samples include the errors made by the fine-tuned code LM, teaching the same model about the code refinement aligns with our final goal that we expect the code LMs to refine the faulty code generated by itself, \ie ``self"-refinement.

\subsubsection{Aggregate the Problem Description, Faulty Generation, and Execution Results as a Joint Prior Condition}
\label{subsubsec:align_multiple_resources}
\begin{wrapfigure}{r}{0.61\linewidth}
    \centering
    \vspace{-5pt}
    \includegraphics[width=\linewidth]{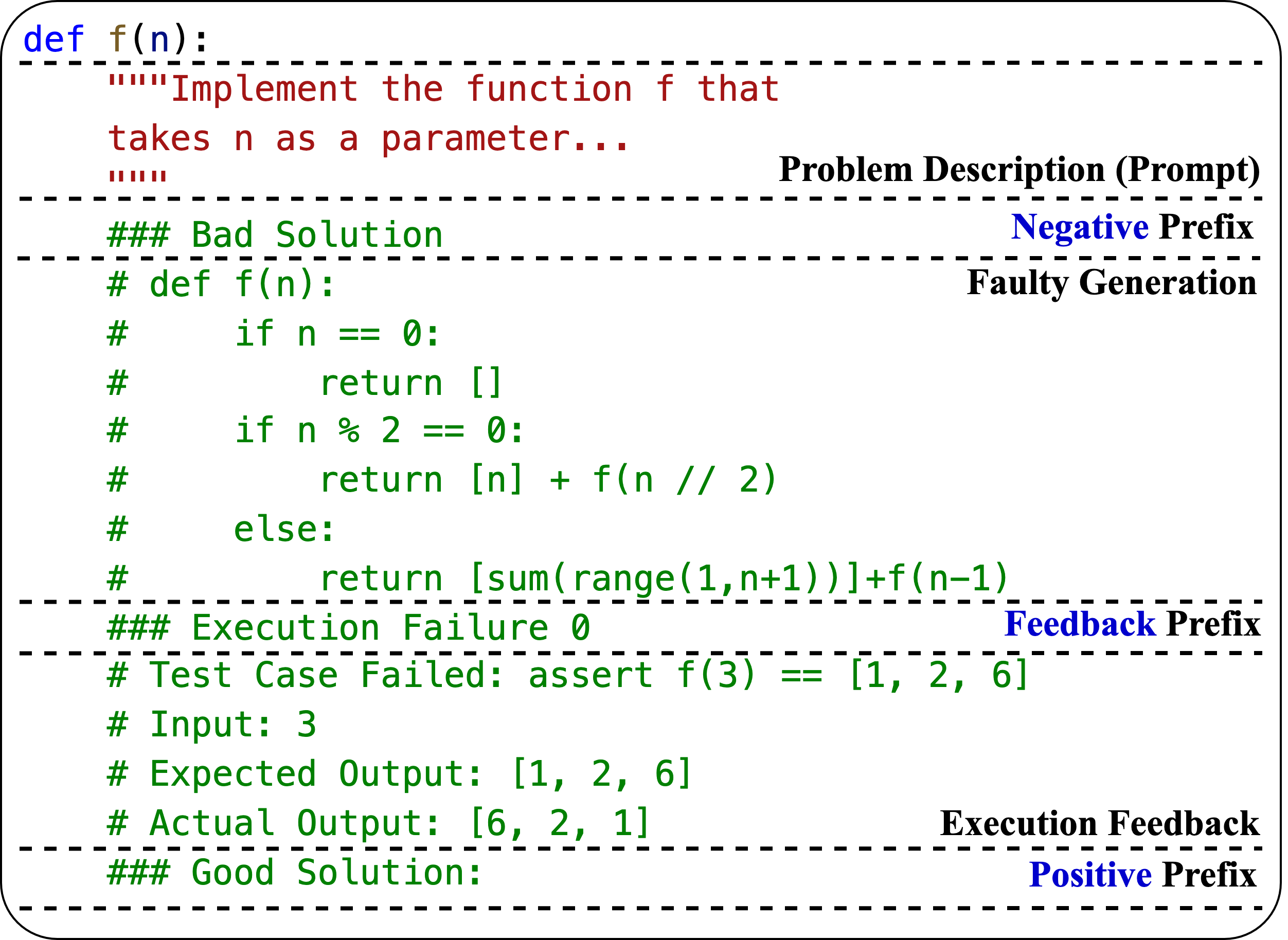}
    \caption{Template to aggregate the problem description, faulty generation, and the execution feedback.}
    \label{fig:aligned_template}
    \vspace{-5pt}
\end{wrapfigure}

To effectively teach code LMs about code refinement, we propose a template that aggregates information from multiple resources. We show the template in Figure~\ref{fig:aligned_template}. To ensure the code's naturalness remains intact, we adopt a strategy of encapsulating our template within docstrings or comments. This template comprises six essential components. 

First, we encapsulate the problem description within docstrings. Next, we introduce a negative prefix that signifies the beginning of a flawed code generation, followed by the actual erroneous code. In a manner similar to the negative prefix, we employ an execution prefix to precede the feedback regarding the code's execution. This feedback meticulously outlines the test results, encompassing specific failed test cases, input data, the anticipated output, and the actual misaligned output. Finally, our template concludes with a positive prefix designed to prompt the generation of the correct code solution.

\subsubsection{Learning to Self-Refine}
\label{subsubsec:learning_to_self_refine}
With the proposed template above, we efficiently aggregate all information together for the model to jointly attend to. To learn the refinement, the model will be fed with the aggregated information and learn to predict the canonical solution of this problem. Specifically, we define the aggregated information, at a high level, as the combination of problem description, $NL$, faulty generation, $FG$, and the execution feedback, $EF$, so $AGGR = \{NL, FG, EF\}$. The target canonical solution, $C$ will be predicted token by token, and thus the loss can be represented as:

\begin{equation}
    \mathcal{L}_{self-refine} = \sum_{n \in |C|} - log P(c_n ~| ~NL, FG, EF, c_1, c_2, ..., c_{n-1})
\end{equation}

Though the learning process is mostly about the next token prediction~\cite{Radford2018gpt1, Radford2019gpt2}, it is quite effective at learning the self-refinement, which will be illustrated in Section~\ref{subsec: main_result}. The main reason for its effectiveness is that the learning objective forces the model to learn different knowledge from distinct resources of information. To correctly predict each code token $c_i$, the model needs to understand $NL$, $FG$, and $EF$ respectively, and jointly decide how to take advantage of each resource.

\noindent\textbf{Past Generation Mask (PGM)} When aiming to refine faulty code generations, there is a risk that, during the training process, the model pays too much attention to the prior faulty generation, \ie $FG$, it might end up taking shortcuts by copying the tokens to minimize the self-refinement loss mentioned above, since the faulty code has significant overlap, in terms of tokens, with the target implementation (Section~\ref{subsubsec:distill_weakness}). Instead of genuinely understanding and rectifying the code based on the feedback, the model could resort to merely copying or reproducing large portions of the past generation. Such shortcuts are difficult to detect during the training, as the loss will not significantly differ regardless of model takes the shortcut or not, but the model relying on the shortcuts becomes useless when being deployed.

To alleviate such risk, we are motivated by the efficient masking approaches in deep learning literature, such as dropout~\citep{srivastava2014dropout} and forgetful causal mask~\citep{liu2023forgetful}, and we propose to randomly mask $p$\% of tokens in the faulty generation $FG$, which is the main resource for copying. Concretely, we manipulate the attention mask in the Transformer architecture~\citep{vaswani2017transformer}, which will make the masked token invisible to other tokens without the explicit removal of tokens. We will study the effectiveness of this masking strategy in Section~\ref{subsec: piece_eval}. Note that the masking ratio, $p$, is a tunable hyperparameter, and we will study its impact in Section~\ref{subsec:ablation}.

\subsubsection{Mixture of data resources} 
\label{subsubsec:mix_of_data}

Data is always the key to the success of language model training, and existing code LMs~\cite{li2023starcoder, rozière2023codellama} have illustrated the necessity of combining multiple data resources with carefully tuned proportions. In the quest to enhance the proficiency of our model for self-refinement, we identified the necessity to curate an optimal mix of self-refine samples, as described in Section~\ref{subsubsec:distill_weakness}, and the canonical solutions used in Section~\ref{subsubsec:approach_ft_w_correct_code}. The rationale behind this strategy is to balance \tool's capability in acceleration and exploration mode. 

The proportion of these data resources also introduces a nuanced layer of complexity to the training regimen. An imbalance, especially an excessive reliance on self-refine samples, could lead the model to overfit code refinement. In such a scenario, it expects, and perhaps becomes overly dependent on, the trifecta of the natural language prompt, the prior faulty generation, and the execution feedback. This could compromise its innate capacity to generate code accurately in situations where only a simple prompt is provided. Conversely, if the original code samples dominate the training set, the model might not sufficiently internalize the mechanisms of self-correction and iterative refinement. This intricate balance underscores the importance of meticulous data engineering in training a versatile and robust code language model. We will study the impact of such data mixture in Section~\ref{subsec:ablation}

\subsection{Phase-III: Iterative Self-Refinement with Execution Feedback}
\label{subsec: approach_inference}

While the Phase-II training is formulated as a one-step refinement from the faulty code to its correct version, we implement the automated inference framework as an iterative programming workflow. There are three main reasons. First, we assume that one-step refinement is not enough to fulfill the model's best capacity, as there is always a gap between the perfect performance that the training tries to approach and the final inference performance that could really be achieved. Second, the diversity of self-refinement samples should enable \tool with the iterative capability of refinement Third, our design is to harmonize the automated code generation with human-like introspection and iterative improvement. By incorporating feedback loops, setting clear stop criteria, and leveraging the structured aggregation from Phase-II, we aim to transform code generation from a one-time event to a dynamic, self-evolving process. This methodology, inspired by human programming practices, sets the stage for more resilient, accurate, and context-aware code outputs.

Specifically, as shown in Figure~\ref{fig:overview}, upon receiving the problem description, the model initially produces what it perceives as the optimal code. This generated code is then automatically tested against the relevant test suite. Should any test cases fail, our framework seamlessly compiles the original description, the incorrect code, and the associated execution feedback, all according to the template from Phase-II. This consolidated input is then presented back to the model, guiding its subsequent refinement efforts. The process of self-refinement is not endless considering the overhead and computation it might cost, and it will be stopped by three scenarios: (1) if the refined code successfully passes the test suite, or (2) if a threshold for the maximum self-refine times is reached, or (3) if the refined code remains the same as the previous faulty code, which means the model could no more improve the code according to the execution feedback.

\section{Experimental Setup}

\subsection{Training Datasets}
\label{subsec:training_datasets}
We formulate the training dataset for \tool with samples collected from CodeContest~\citep{li2022alphacode}. CodeContest includes programming competition problems curated from Codeforces~\citep{codeforces2020} and CodeNet~\citep{puri2021codenet}. Each problem is accompanied by both canonical solutions and wrong human submissions written in three programming languages (Python, Java, C++), as well as executable test cases to verify the correctness of solutions.

To construct our training datasets, we focus on the training split of CodeContest which originally contained 13,328 problems. We further filter out those problems with long descriptions (> 512 tokens), as they could take too much input length of Transformer-based LM and leave insufficient length to include the code solutions that \tool will be trained to predict. In addition, we remove the problems without Python solutions. Finally, our training set ended up with 7,108 problems from CodeContest. We also constructed a held-out validation set from CodeContest valid split, which originally contained 117 problems, to monitor the training process and the model's generalizability to unseen data during training. After the same filtering as our training set, we keep 77 problems in the validation set. 

To fine-tune the code LMs for the data preparation phase (Section~\ref{subsubsec:approach_ft_w_correct_code}), we sample up to 50 canonical solutions for each of the problems, resulting in 233,703 training samples and 3,833 validation samples. To build the samples for the self-refinement training (Section~\ref{subsubsec:distill_weakness}), we sample 10 generations for each problem from the fine-tuned code LMs and execute the test suites to verify their correctness and collect the execution feedback. In total, we have 71,080 training samples and 770 validation samples for the self-refinement learning phase.

\subsection{Evaluation Benchmarks}
To evaluate the code generation performance, we use three popular programming benchmarks: HumanEval~\citep{chen2021codex}, MBPP-S~\citep{austin2021synthesis}, and APPS~\citep{hendrycks2021apps}.

\noindent\textbf{HumanEval}~~ HumanEval contains 164 hand-written programming problems by \citet{chen2021codex}. Each problem includes a function signature and docstring as the description, and the model is asked to predict the function body. Each problem also maintains on average 7.7 unit test cases, in the format of assertions, to verify the correctness of the model prediction.

\noindent\textbf{MBPP-S}~~ Mostly Basic Programming Problems (MBPP) benchmark contains 974 short Python programs constructed by crowd-sourcing a pool of crowdworkers who only have basic knowledge of Python. Each problem contains a short problem statement and provides 3.0 unit test cases to verify the functionality. \citet{austin2021synthesis} further sanitized the benchmark by manually inspecting and removing the ambiguous or unexpected problems, resulting in a version called MBPP-Sanitized (or MBPP-S for short) with a total of 426 problems of better quality. In our evaluation, we apply this sanitized version of MBPP-S for more accurate evaluation.

\noindent\textbf{APPS}~~ APPS is a benchmark collected to evaluate the code generation models. It includes a total of 10,000 problems collected from open-access coding websites, accompanied by a total of 131,777 test cases and 232,421 canonical solutions. To avoid data leakage, we focus on the test split with 5,000 problems. In addition, given that APPS has significantly overlapped data resources with our training data from CodeContest, we apply an aggressive filtering strategy to deduplicate. Specifically, we exhaustively calculate the fuzzy edit similarity between the problem description in APPS and in CodeContest, and if the similarity is over 60\%, we will remove the whole problem from APPS to avoid the data memorization issue. Finally, after the filters, we ended up with 1,280 problems as our evaluation problem set.

\subsection{Models}
To illustrate the generalizability of \tool's design, we train four variants of \tool with varied sizes of model parameters: 350M, 1B, 2.7B, and 3B. These variants are initialized from the checkpoints of two open-source code LMs families: CodeGen~\citep{nijkamp2023codegen} and StarCoder~\citep{li2023starcoder}. 

\noindent\textbf{CodeGen} is a set of autoregressive code LMs, with varied sizes (350M to 16B), pre-trained using next-token prediction objective on a large-scale dataset collected from \textsc{ThePile}~\citep{gao2021thepile}, Google \textsc{BigQuery}, and \textsc{BigPython}~\citep{nijkamp2023codegen}. The dataset includes both natural language and code samples with over 550 billion tokens. The model maintains the context length of 2,048 BPE~\cite{kudo-richardson-2018-sentencepiece} tokens.

\noindent\textbf{StarCoder} is a set of code LMs, with varied sizes (1B to 15.5B), that are pre-trained on over 1 trillion tokens from The Stack~\cite{kocetkov2022stack} dataset, using both next-token prediction and the fill-in-the-middle~\citep{fried2023incoder, bavarian2022efficient} objectives. StarCoder family could consume up to 8,192 BPE tokens.

Specifically, \tool variants are initialized from CodeGen-350M, StarCoder-1B, CodeGen-2.7B, and StarCoder-3B respectively. We load the checkpoints from the Hugging Face Model Hub~\citep{hfmodelhub}.

\subsection{Configurations and Hyperparameters}
\label{subsec:config_param}
We conduct our experiments on 2$\times$ NVIDIA GeForce RTX 3090 with 24GB GPU memory each. The model is implemented mainly with PyTorch~\citep{paszke2019pytorch} and Hugging Face Transformers ~\citep{wolf-etal-2020-huggingface} library.

For training, we consider a batch size of 512 samples with 2,048 BPE~\citep{kudo-richardson-2018-sentencepiece} tokens. We apply a standard learning rate descending strategy of code LM~\citep{nijkamp2023codegen, li2023starcoder, rozière2023codellama} that the early phase of training applies a higher learning rate than the later phase, and small models use a higher learning rate than large models. Concretely, for the fine-tuning in the data preparation phase (Section~\ref{subsec:data_preparation}), \tool uses [5e-5, 2e-5, 1e-5, 1e-5] for [350M, 1B, 2.7B, 3B] respectively, and for the self-refinement learning phase (Section~\ref{subsec:self_refine_training}), \tool uses [2e-5, 2e-5, 5e-6, 5e-6] for the aforementioned model sizes respectively. All the training applies a cosine learning rate decay scheduler with warmup steps. For both the fine-tuning in the data preparation phase and the self-refinement learning, we train \tool for only one epoch on the corresponding dataset. The PGM masking rate (Section~\ref{subsubsec:learning_to_self_refine}) for training is 0.05, and the ratio of self-refinement samples is 25\% (Section~\ref{subsubsec:mix_of_data}).

For inference, we adapt the standard nucleus sampling~\citep{Holtzman2020nucleussampling} with the top-p probability of 0.95. For HumanEval and MBPP-S benchmarks, we ask the model to generate up to 256 BPE tokens, and for the APPS benchmark, the model will generate up to 512 BPE tokens, as the latter's problem is more difficult and the solutions are typically longer. For the self-refinement process during the inference, we set up a maximal refinement step of 4, and this choice is based on the tradeoff between the inference overhead and the performance. We will study more about the number of refinement steps in Section~\ref{subsubsec:self_refine_continue_improve_cycle} and the inference overhead in Section~\ref{subsubsec:overhead_study}

\section{Evaluation}
In this section, we evaluate \tool and analyze it by asking the following research questions: 

\begin{itemize}
    \item RQ1: How effective is \tool in code generation compared to the existing code LMs?
    \item RQ2: How do \tool's different designs contribute to its performance?
    \item RQ3: How is \tool's iterative self-refinement different from Top-K generations?
    \item RQ4: How will PGM's masking ratio and data mixture proportion affect the performance?
\end{itemize}

\subsection{RQ1. \tool's Performance in Code Generation}
\label{subsec: main_result}

In this section, we present the main evaluation regarding \tool's performance in code generation, and we illustrate its effectiveness by comparing it to existing code LMs of varied size, across three popular programming benchmarks. 

\noindent\textbf{Setup.} We evaluate the model's code generation capability in two main settings: one-time generation and iterative self-refinement. For the one-time generation, we follow the original design of the evaluation benchmarks~\citep{chen2021codex, austin2021synthesis, hendrycks2021apps}, where the description of programming challenges will be fed into the model as the prompt, and the model will implement the program accordingly. We use the accompanied test suites of each programming challenge to verify the correctness of the prediction, where a test suite includes multiple test cases to evaluate the code with distinct perspectives.

For the iterative self-refinement performance evaluation, we follow the workflow of \tool's inference framework (Section~\ref{subsec: approach_inference}). The model will take the one-time generation as the starting point for the refinement, and repeat the process up to four times (Section~\ref{subsec:config_param}). If the generated code passes the test suite at a time step, the self-refinement of this sample will be terminated, and this sample will be regarded as a success. If the model cannot predict a correct program with four times of self-refinement, the sample will be regarded as a failure.

\noindent\textbf{Baselines.} We consider vanilla open-source code LMs from CodeGen~\citep{nijkamp2023codegen} and StarCoder~\citep{li2023starcoder} families as the first set of baseline. Specifically, we consider CodeGen-350M, StarCoder-1B, CodeGen-2.7B, and StarCoder-3B. CodeGen and StarCoder are two of the most popular code LMs with state-of-the-art code generation capabilities, and since \tool variants are initialized from these models to continue learning the self-refinement (Section~\ref{subsubsec:approach_ft_w_correct_code}), comparing with them will directly reflect the effectiveness of \tool's proposed training.

In addition, we create a stronger set of baselines from the fine-tuned checkpoints we introduced in Section~\ref{subsubsec:approach_ft_w_correct_code}. We further train these checkpoints with the canonical solution from \tool's training datasets but no self-refinement signals are included. This set of baselines has two main improvements compared to the vanilla code LMs. First, it is trained with the same data samples with the same amount of epochs as \tool, which directly verifies whether the naive training on the canonical solution is enough to grant the model self-refine capability. Second, this set of baseline is trained with canonical solutions only, so comparing to it illustrates the value of \tool's self-refinement training that learns to jointly understand the faulty generation and execution feedback as well. We call this set of baselines ``code LMs fine-tuned with correct code".

Note that, since baseline models have never been exposed to the template we designed for \tool to perform self-refinement (Section~\ref{subsubsec:align_multiple_resources}), we notice that applying such a template to baseline models will hurt their performance by misleading the model to keep generating comments rather than the real code. Alternatively, to maximize their performance for a fair comparison, we borrow the idea from existing work~\cite{chen2023teaching} to wrap the faulty code and the execution feedback into the docstring as plain text, and it turns out that the baseline models could normally generate code after this adaption.

\begin{table*}[!ht]
\centering
\caption{Comparing \tool's performance with baseline models in both one-time generation and iterative self-refinement settings. We consider four sizes of code LMs to discuss, ranging from 350M to 3B. The first row of each section is the ``vanilla code LM" baseline, the second row of each section is the ``code LMs fine-tuned with correct code" baseline, and the last row is \tool variant of the same size.}
\label{tab:rq1_result}
\begin{tabular}{l c c c c c c}
\toprule

\multirow{3}{*}{\textbf{Model}}  &\multicolumn{3}{c}{\textbf{One-time}}&\multicolumn{3}{c}{\textbf{Self-Refine}}\\\cmidrule(lr){2-4} \cmidrule(lr){5-7} 

&\textbf{ HumanEval} &\textbf{ MBPP-S} & \textbf{APPS} & \textbf{HumanEval} & \textbf{MBPP-S} & \textbf{APPS}\\\hline\hline

CodeGen--350M & \colorgray 12.2 & \colorgray 19.0 & \colorgray 6.9 &\coloryellow 12.2~~\blimprove{+0.0\%} & \coloryellow 21.8~~\blimprove{+14.8\%} & \coloryellow 6.9~~\blimprove{+0.0\%} \\

\quad + FT w/ Correct & \colorgray 12.2 & \colorgray 19.2 & \colorgray \textbf{7.5} &\coloryellow 14.0~\blimprove{+14.9\%} & \coloryellow 23.0~\blimprove{+19.5\%} & \coloryellow 7.7~\blimprove{+2.8\%}\\\hdashline%

\tool--350M & \colorgray \textbf{14.0} & \colorgray \textbf{19.9} & \colorgray \textbf{7.5} &\coloryellow \textbf{20.7}~\improve{\textbf{+47.9\%}} & \coloryellow \textbf{32.6}~\improve{\textbf{+63.5\%}} & \coloryellow \textbf{8.7}~\improve{\textbf{+15.6\%}} \\\hline\hline

StarCoder--1B & \colorgray 15.9 & \colorgray 25.8 & \colorgray 7.3 &\coloryellow 16.5~~\blimprove{+3.8\%} & \coloryellow 28.1~~\blimprove{+9.1\%} & \coloryellow 7.3~~\blimprove{+0.0\%} \\

\quad + FT w/ Correct & \colorgray \textbf{18.3} & \colorgray \textbf{26.0} & \colorgray 8.6 &\coloryellow 18.9~~\blimprove{+3.3\%} & \coloryellow 28.3~~\blimprove{+9.0\%} & \coloryellow 9.3~~\blimprove{+8.3\%} \\\hdashline%

\tool--1B & \colorgray \textbf{18.3} & \colorgray 25.8 & \colorgray \textbf{8.9} &\coloryellow \textbf{22.0}~~\improve{\textbf{+20.0\%}} & \coloryellow \textbf{35.8}~~\improve{\textbf{+39.1\%}} & \coloryellow \textbf{10.9}~\improve{\textbf{+22.8\%}}~~ \\\hline\hline

CodeGen--2.7B & \colorgray \textbf{21.9} & \colorgray 34.7 & \colorgray 7.1 &\coloryellow 23.8~~\blimprove{+8.4\%} & \coloryellow 35.4~~\blimprove{+2.0\%} & \coloryellow 7.1~~\blimprove{+0.0\%} \\

\quad + FT w/ Correct & \colorgray \textbf{21.9} & \colorgray \textbf{36.8} & \colorgray 9.0 &\coloryellow 23.8~~\blimprove{+8.4\%} & \coloryellow 40.5~~\blimprove{+10.2\%} & \coloryellow 9.7~~\blimprove{+7.9\%} \\\hdashline%

\tool--2.7B & \colorgray 21.4 & \colorgray 35.8 & \colorgray \textbf{9.1} & \coloryellow \textbf{29.3}~~\improve{\textbf{+37.1\%}} & \coloryellow \textbf{48.5}~~\improve{\textbf{+35.3\%}} & \coloryellow \textbf{11.6}~~\improve{\textbf{+27.6\%}} \\\hline\hline

StarCoder--3B & \colorgray 23.8 & \colorgray 35.1 & \colorgray 7.3 &\coloryellow 26.8~~\blimprove{+12.8\%} & \coloryellow 40.5~~\blimprove{+15.3\%} & \coloryellow 7.4~~\blimprove{+1.1\%} \\

\quad + FT w/ Correct & \colorgray \textbf{24.4} & \colorgray 35.8 & \colorgray \textbf{9.0} &\coloryellow 24.4~~\blimprove{+0.0\%} & \coloryellow 40.8~~\blimprove{+13.7\%} & \coloryellow 10.2~~\blimprove{+13.9\%}  \\\hdashline%

\tool--3B &\colorgray \textbf{24.4} & \colorgray \textbf{36.3} & \colorgray \textbf{9.0} &\coloryellow \textbf{29.9}~~\improve{\textbf{+22.5\%}} & \coloryellow \textbf{51.3}~~\improve{\textbf{+41.3\%}} & \coloryellow \textbf{11.3}~~\improve{\textbf{+25.3\%}} \\\hline\hline
\end{tabular}
\end{table*}

\noindent\textbf{Findings.} The results of one-time code generation and self-refinement across four sizes of code LMs (350M, 1B, 2.7B, 3B) are reported in Table~\ref{tab:rq1_result}.

\paragraph{Finding-1.1: Self-Refinement Capacity Does Not Come Along Naturally with Code LMs' Pre-training.}\footnote{Note that our finding is concluded based on the code LMs with sizes between 350M and 3B parameters. We did not study the emergent ability of significantly larger models (\eg with 175B and 540B parameters)~\citep{wei2022emergent}.} As we can see from the table, baseline code LMs, though pre-trained on tons of code samples, primarily focus on one-time code generation and struggle to self-refine the faulty generation in the past by understanding the execution feedback. It is also evident that, though the one-time code generation performance aligns with the scaling law of neural language models~\citep{kaplan2020scaling}, the increase in model sizes does not necessarily translate to better self-refinement capabilities. For example, baseline code LMs with sizes ranging from 350M to 2.7B all fail to correctly self-refine a single example in APPS (0.0\% improvement after performing self-refinement). This verifies our assumption that existing code LMs are not exposed to sufficient signals during the pre-training to learn how to self-refine, and naively stacking more model parameters or collecting more data from the wild is not that helpful.

We notice that fine-tuning the baseline code LMs with only semantically correct code, as indicated by the "+ FT w/ Correct" rows, shows only marginal improvements in both one-time generation and self-refinement. This means that merely training on correct code is not enough to equip the models with the skill of rectifying their own mistakes, which requires a decent understanding of the execution feedback and the capability of fixing the errors accordingly.

\paragraph{Finding-1.2: \tool is Effective at Improving Code LM's Self-Refinement Capacity and Generalizable to Varied Model Sizes.} As shown in Table~\ref{tab:rq1_result}, across three programming benchmarks, \tool consistently levels up the code generation performance with self-refinement, resulting in up to 63.5\% relative improvement compared to one-time generation. \tool with 350M parameters notably outperforms StarCoder-1B, which maintains 3$\times$ more parameters, across all three benchmarks, highlighting the effectiveness of \tool's self-refinement and the efficiency of the proposed training strategy.

Compared to baseline models, \tool has notably stronger capabilities at correcting faulty generations in the past. For example, while the vanilla CodeGen-2.7B model could not refine its own wrong predictions in APPS, while \tool-2.7B successfully refines 27.6\% more samples which it failed to predict for the first time. This suggests that, by training to refine code based on execution feedback and previous mistakes, CYCLE builds a more holistic understanding of code, enabling it to not just generate code, but also understand its intricacies, potential pitfalls, and nuances.

In addition, as we introduced in Section~\ref{sec:approach}, \tool does not require to be trained from scratch, and rather, it loads pre-trained code LMs as the starting point and further teaches the model to self-refine. As we can see from Table~\ref{tab:rq1_result}, \tool is effective at varied sizes of code LMs and consistently boost the performance for two different neural architectures\footnote{At a high level, CodeGen and StarCoder are both GPT-like Transformer decoder, but their concrete neural architectures (\eg positional embedding and attention layers) are not the same. More details can be referred from: StarCoder (\url{https://github.com/huggingface/transformers/blob/main/src/transformers/models/gpt_bigcode/modeling_gpt_bigcode.py}) and CodeGen (\url{https://github.com/huggingface/transformers/blob/main/src/transformers/models/codegen/modeling_codegen.py)}}, CodeGen and StarCoder. This empirically proves the generalizability of \tool that it can always be applied in a plug-and-play style, suggesting the potential of taking advantage of more powerful code LMs that will be trained and released in the future.

\paragraph{Finding-1.3: \tool Maintains Decent Capacity in One-time Code Generation.} One recognized risk of continuing training unsupervised/self-supervised models, which are pre-trained with large-scale data, on limited, carefully crafted samples is the \emph{model shift} problem~\citep{wang2021selftuning, wang2015generalization}. The model could shift towards the local optimal of the limited but new data while the previous knowledge is eventually wiped out. As we can see in Table~\ref{tab:rq1_result}, \tool does not suffer this issue; it successfully maintains the one-time generation capability. This empirically verifies that our proposed alignment of prompt, fault generation, and execution feedback (Section~\ref{subsubsec:align_multiple_resources}) is intuitive to the model, preventing the significant distribution shift. Also, the data mixture strategy (Section~\ref{subsubsec:mix_of_data}) plays a role in neutralizing the distribution gap, and we will further analyze this strategy in Section~\ref{subsec:ablation}. Interestingly, the 350M, 2.7B, and 3B versions of \tool perform slightly better than the standard baseline code LM. We believe the primary reason for this is the way \tool is trained. During its training process, \tool is exposed to both faulty and correct code, while our specialized training method encourages the model to only generate the correct one(Section~\ref{subsubsec:learning_to_self_refine}). This gives \tool a unique advantage: it understands what good and bad code look like respectively, but it's specifically trained to produce the good one. In comparison, baseline code LMs are not exposed to such preference signals during their pre-training.

\begin{finding}
While maintaining the one-time code generation capacity, mostly improving marginally, \tool significantly boosts the code LMs' self-refine capacity by learning to understand the execution feedback and faulty code generated in the past. \tool enables existing code LMs to match or beat larger models with 3$\times$ more parameters.
\end{finding}

\subsection{RQ2. Impacts of \tool's Different Designs on Code Generation}
\label{subsec: piece_eval}

While we have concluded that \tool's proposed approach, as a whole, is effective at code generation in Section~\ref{subsec: main_result}, now we delve deeper to analyze how the isolated design contributes to the performance individually. 

\subsubsection{Self-Refinement Continuously Improves \tool's Performance}
\label{subsubsec:self_refine_continue_improve_cycle}
\begin{figure}[!h]
  \centering
  \begin{subfigure}[b]{0.32\textwidth}
    \includegraphics[width=\textwidth]{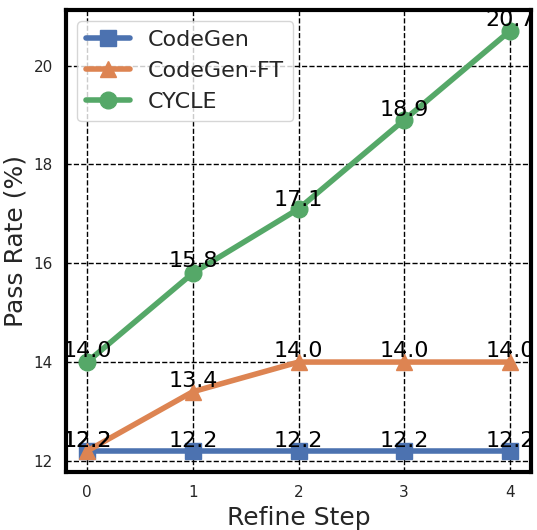}
    \caption{HumanEval}
    \label{fig:self_refine_humaneval}
  \end{subfigure}
  \begin{subfigure}[b]{0.32\textwidth}
    \includegraphics[width=\textwidth]{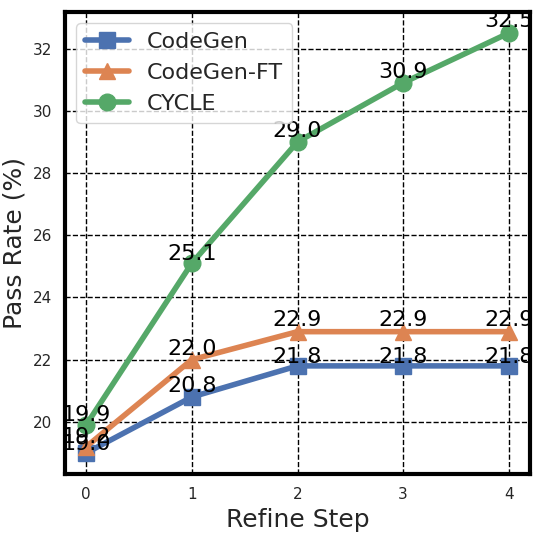}
    \caption{MBPP-S}
    \label{fig:self_refine_mbpp}
  \end{subfigure}
  \begin{subfigure}[b]{0.32\textwidth}
    \includegraphics[width=\textwidth]{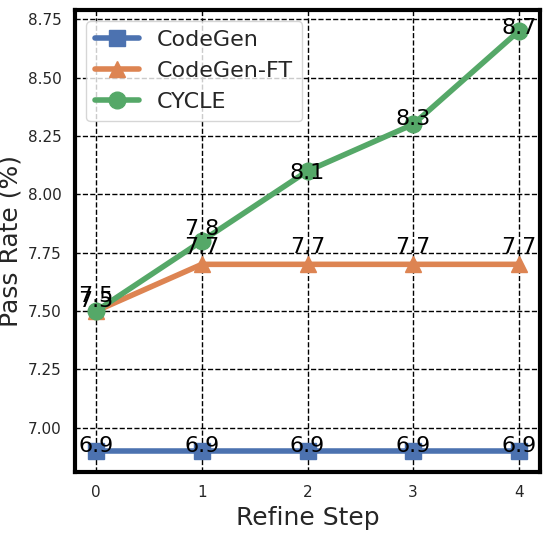}
    \caption{APPS}
    \label{fig:self_refine_apps}
  \end{subfigure}
  \caption{Performance improvement with self-refinement. The blue curve represents the ``vanilla code LM baseline", using CodeGen-350M. The orange curve represents the ``code LMs fine-tuned with the correct code". The green curve represents \tool-350M.}
  \label{fig:self_refine_impact}
  \vspace{2mm}
\end{figure}

Teaching code LMs to self-refine is the core idea behind \tool's design, so we first study how self-refinement boosts \tool's code generation capability. To do this, we plot the \tool-350M's performance at each refinement step, in Figure~\ref{fig:self_refine_impact}, to reveal the trend. We also plot the baseline models' trends as a comparison. 

We can see that \tool's performance is improved with each refinement step across all three benchmarks. This reveals that \tool is able to eventually correct its own error, step by step, by understanding the mismatch between the expected dynamic behavior and real implementations from the execution feedback. In addition, the figure reveals that \tool still does not reach its best performance after four times of refinements, as the curve has not plateaued yet, highlighting its potential to achieve much better results. \tool's ability to self-refine also aligns better with human developers' iterative programming practice, where they learn from mistakes and continuously improve. \tool exemplifies how code LMs can be designed to imitate iterative programming, bringing real-time improvement to developers' code interactively. 

In contrast, baseline models, including the one fine-tuned with the same data as \tool, typically plateau after one or two steps of refinement. For example, in Figure~\ref{fig:self_refine_mbpp} that evaluated with MBPP-S benchmark, \tool and baseline models start at similar performance, but refinement could not bring further improvement after two steps, while \tool continues to go up. This gap sheds light on the necessity of evaluating code LM's capacity for self-refinement: models with similar one-time generation performance might not be similarly helpful, as the one with better self-refinement capacity could assist developers with iterative programming more efficiently in the realistic deployment.

\subsubsection{Execution Feedback Guides \tool for Better Self-Refinement} Second, we analyze whether the execution feedback helps \tool as expected. To do so, we remove the execution results before each time of refinement, only leaving the problem description, \ie natural-language prompt, and the past, faulty generation as references. The results are shown in Table~\ref{tab:exec_removal}.

\begin{table*}[!ht]
\centering
\caption{\tool's performance decreases when the execution feedback is removed during the self-refinement, while the baseline models could not effectively perceive the execution results.}
\label{tab:exec_removal}
\begin{tabular}{l l l l}
\toprule

\textbf{Benchmark} &\textbf{HumanEval} &\textbf{MBPP-S} & \textbf{APPS} \\\hline\hline

\grayline \multicolumn{4}{c}{CodeGen}\\

w/o exec feedback & 12.8 & 24.1 & 6.9 \\\hline

w/ exec feedback & 12.2~\decrease{\downarrow \textbf{4.7\%}} & 21.8~\decrease{\downarrow \textbf{9.5\%}} & 6.9~\decrease{- \textbf{0.0\%}} \\\hline\hline

\grayline \multicolumn{4}{c}{CodeGen + FT w/ Correct}\\

w/o exec feedback & 14.0 & 26.0 & 8.3 \\\hline

w/ exec feedback & 14.0~\decrease{- \textbf{0.0\%}} & 23.0~\decrease{\downarrow \textbf{11.5\%}} & 7.7~\decrease{\downarrow \textbf{7.2\%}} \\\hline\hline

\grayline \multicolumn{4}{c}{\tool (Ours)}\\

w/o exec feedback & 15.9 & 29.5 & 8.5 \\\hline

w/ exec feedback & \textbf{20.7}\increase{\uparrow \textbf{20.2\%}} & \textbf{32.6}\increase{\uparrow \textbf{10.5\%}}  & \textbf{8.7} \increase{\uparrow \textbf{2.4\%}} \\\hline\hline

\end{tabular}
\end{table*}

It is evident that execution feedback plays an important role in \tool's self-refinement, and removing it significantly drops \tool's performance across all three benchmarks. In contrast, baseline models are not sensitive to this removal. After removing the execution results, baseline models perform comparably or even better, suggesting that these models regard execution results mostly as redundant information or even noise. This highlights existing code LM's weaknesses in understanding execution effects and taking advantage of them.

\subsubsection{Past Generation Mask (PGM) Effectively Prevents the Exact Copy and Improves \tool's Performance} As we introduced in Section~\ref{subsubsec:learning_to_self_refine}, we design PGM to avoid the model taking shortcuts by naively copy-pasting the faulty generation as its prediction, which might also decrease the training loss due to the token overlaps between the faulty code and its refined version~\citep{ding2020patching}. To verify the effectiveness of PGM, we train \tool-350M-w/o-PGM disabling the PGM (\ie setting the masking ratio to be 0\%) and compare its performance with the standard \tool-350M. 

We studied three perspectives of their differences. First, we compare their overall performance of code generation, \ie the test suite pass rate as we reported in Section~\ref{subsec: main_result}. Second, we analyze the edit distance, in terms of code tokens, between the faulty generation and its refined code predicted by \tool-350M-w/o-PGM and \tool, where a higher value indicates the model modifies more tokens to correct the fault. Third, we study how many exact copies happen along the way of the self-refinement process until it reaches the maximum refine limit. A higher exact copy rate indicates the model copy-pastes the fault generation as the self-refinement prediction more frequently.

\begin{table*}[!ht]
\centering
\caption{Impact of PGM on \tool's performance.}
\label{tab:pgm_study}
\resizebox{1\textwidth}{!}{
\begin{tabular}{l c c c c c c c c c}
\toprule

\multirow{3}{*}{\textbf{\tool}}  &\multicolumn{3}{c}{\textbf{Pass Rate \%~(\textbf{$\uparrow$})}}&\multicolumn{3}{c}{\textbf{Token Edit Distance ($\uparrow$)}} &\multicolumn{3}{c}{\textbf{Exact Copy Rate \%~(\textbf{$\downarrow$}})}\\\cmidrule(lr){2-4} \cmidrule(lr){5-7} \cmidrule(lr){8-10} 

&\textbf{ HEval} &\textbf{ MBPP-S} & \textbf{APPS} & \textbf{HEval} & \textbf{MBPP-S} & \textbf{APPS} & \textbf{HEval} & \textbf{MBPP-S} & \textbf{APPS}\\\hline\hline

w/o PGM & 20.12 & 31.38 & 7.81 & 42.44 & 31.34 & 162.67 & 8.23 & 15.66 & 1.24 \\\hdashline%

w/ PGM &\textbf{20.73} & \textbf{32.55} & \textbf{8.67} &\textbf{45.12} & \textbf{33.17} & \textbf{162.72} &\textbf{7.12} & \textbf{12.86} & \textbf{1.21}\\\hline\hline

\end{tabular}}
\end{table*}

The comparison is shown in Table~\ref{tab:pgm_study}. It is clear that PGM helps to improve the overall performance of self-refinement across all three benchmarks. Also, when PGM is enabled, the model tends to edit more tokens, on average, to refine the faulty generation, and avoid the exact copy at its best.

\begin{finding}
The core feature, self-refinement, significantly enhances \tool's capability, aligning with iterative programming practices in human developers. In addition, the execution feedback is vital for \tool's self-improvement, with its removal leading to a notable drop in performance. Also, Past Generation Mask (PGM) effectively prevents naive copy-pasting, leading to better overall performance and more diverse code edits for refinement.
\end{finding}

\subsection{RQ3. Relationship Between \tool's Self-Refinement and Top-K Generation}
\label{subsec: topk_discussion}

To more comprehensively evaluate code LMs' capacity in code generation, \citet{chen2021codex} proposes to generate up to K sequences simultaneously to explore a more diverse search space and use test cases to pick the correct ones as the final prediction. Later, open-source code LMs~\citep{ nijkamp2023codegen, li2023starcoder, rozière2023codellama} incorporate such a setting as an additional evaluation for the code generation task. 

In this section, we explain that \tool's self-refinement for code generation is an orthogonal direction and maintains comparable overhead to the top-k generation. In addition, the self-refinement is applicable to the top-k generations to further improve the overall performance.

\subsubsection{Preliminary.} When generating code autoregressively, code LMs estimate the probability of each possible next token based on the given context. While it is intuitive to always choose the most probable next token during the generation, which is known as greedy decoding, the single generation might not be able to fully expose the diverse knowledge that the model has learned. To explore a wider search space while ensuring coherence, several techniques~\citep{Holtzman2020nucleussampling, tillmann-ney-2003-beam-search} are applied to code LMs for generating multiple predictions simultaneously. Existing code LMs~\citep{chen2021codex, nijkamp2023codegen, li2023starcoder, rozière2023codellama} have shown that such a top-k generation could improve the overall accuracy.

\noindent\textbf{Beam Search.}~~Beam search is a deterministic search algorithm for sequence generation, extending greedy decoding to generate multiple, most probable sequences. It explores multiple candidate sequences simultaneously by retaining the top-k most promising ones at each step, and the generated sequences so far will be ranked based on the cumulative token probabilities. This process is repeated, gradually expanding the sequences, and at the end, the top-k sequences with the highest likelihood score will be the final output.

While it allows the model to output more than one sequence, it also introduces significantly more computations and inference overhead. For example, maintaining multiple sequences on GPU during generation requires linearly more GPU memories. Also, the repetitive ranking during the generation makes it more time and computation-consuming than a single generation.

\noindent\textbf{Nucleus Sampling with Temperature.}~~\citet{Holtzman2020nucleussampling} propose nucleus sampling, also known as top-p sampling, to enhance the creativity of sequence generation with language models. Instead of rigidly choosing the top-k most probable sequences like beam search, nucleus sampling dynamically selects a subset of the most probable words at each generation step and samples one out of them. 

Specifically, when generating the next token, nucleus sampling first cumulates a probability mass of the most likely tokens, called the \emph{nucleus}, until it exceeds the threshold ``p" (\eg, 0.9). Then a concrete token will be randomly sampled from this dynamically determined probability mass. Nucleus sampling can also be coupled with a temperature parameter, which re-scales the likelihood distribution of words and consequently affects the dynamic selection of the \emph{nucleus}. A higher temperature (\eg 0.8) makes the distribution flatter, involving more tokens into the \emph{nucleus}, while a lower temperature (\eg 0.2) sharpens the distribution, favoring high-probability tokens.

To generate ``k" sequences with nucleus sampling, the typical approach is to duplicate the input ``k" times and apply the sampling independently, where, at each step, the randomness of sampling will make the difference. Similar to beam search, as it maintains multiple sequences on GPU, the extra memory overhead is applied.

For the rest of the section, we conduct experiments to study the code LM's performance with top-k generation using beam search and nucleus sampling with different temperatures and compare the results with \tool's self-refinement, highlighting their difference and the potential interaction. To save the computation, all experiments are conducted with \tool-350M, but the experiments are designed to be generalizable to all sizes of code LMs, and we expect the observations to be maintained among different model sizes.

\subsubsection{\tool's self-refinement is orthogonal to top-k generation} 
\label{subsubsec:top_k_is_orthogonal}
\begin{figure}[!h]
  \centering
  \begin{subfigure}[b]{0.4\textwidth}
    \includegraphics[width=\textwidth]{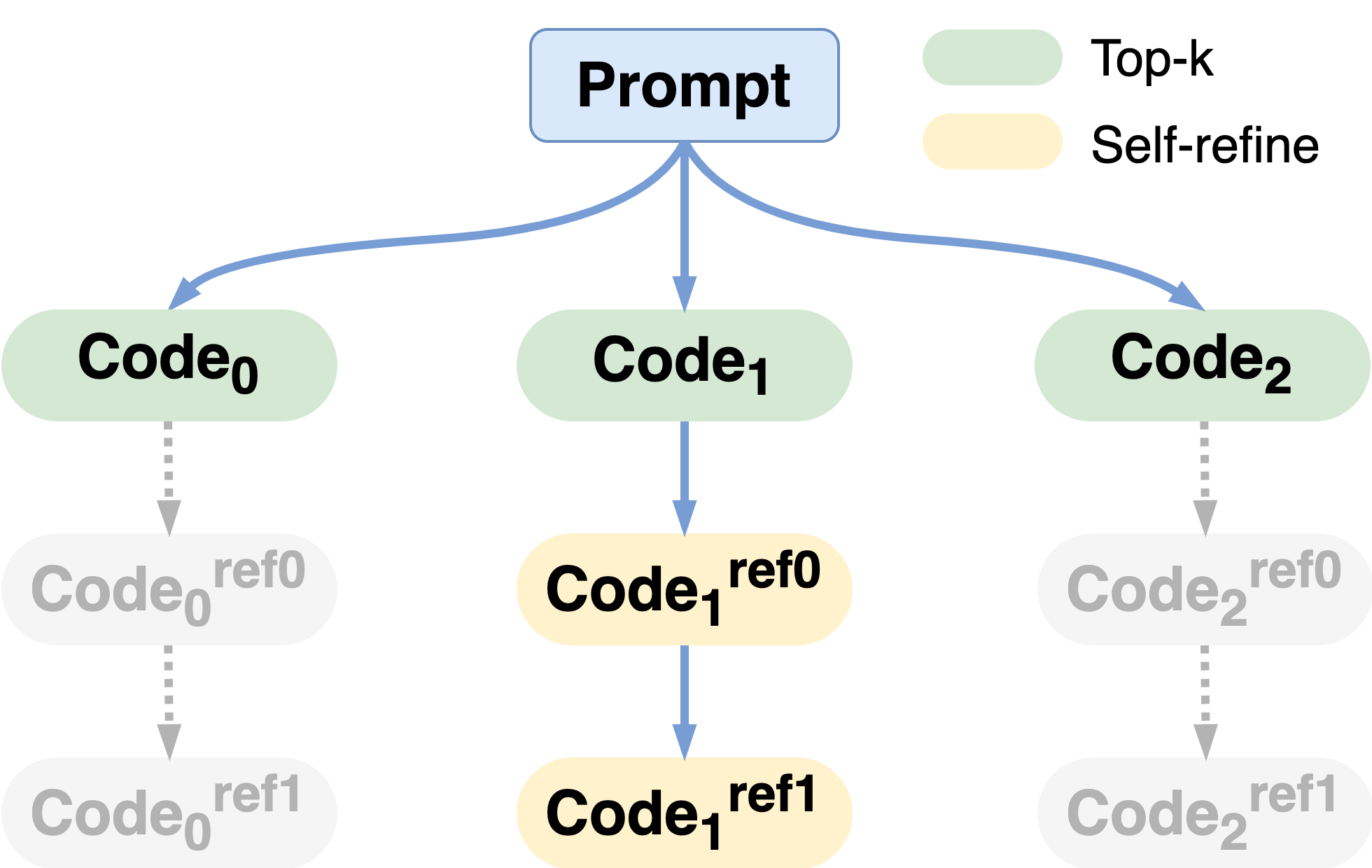}
    \vspace{6mm}
    \caption{Conceptual illustration of the difference}
    \label{fig:top-k_illustration}
  \end{subfigure}
  \hspace{10mm}
  \begin{subfigure}[b]{0.4\textwidth}
    \includegraphics[width=\textwidth]{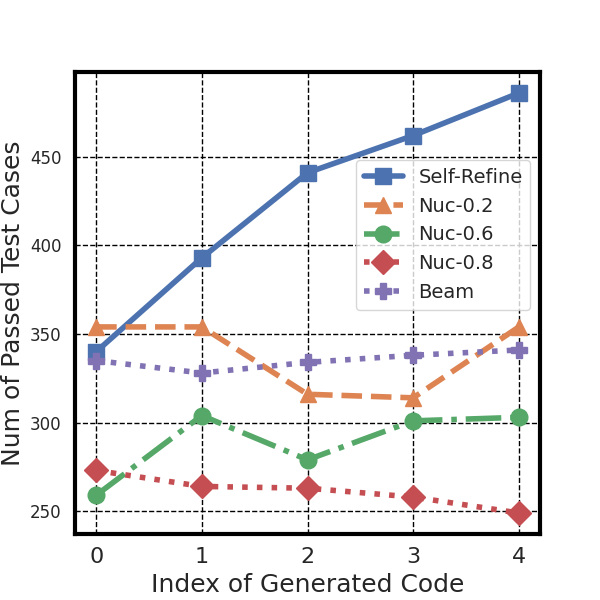}
    \caption{Empirical evidence from MBPP-S}
    \label{fig:top_k_test_case_pass}
  \end{subfigure}
  \caption{\tool's self-refinement is orthogonal to the top-k generation. Conceptually, top-k generation explores in breadth, while self-refinement improves a specific generation in depth. Empirically, when compared to nucleus sampling (with the temperature of 0.2, 0.6, 0.8) and beam search, self-refinement optimizes the generated code towards execution guidance, while top-k produces more diverse programs that pass a similar amount but complementary test cases.}
  \label{fig:top_k_and_self_refine}
  \vspace{2mm}
\end{figure}

As we show in Figure~\ref{fig:top-k_illustration}, the iterative process of self-refinement is conceptually different from the top-k generation. Concretely, top-k generation produces multiple code candidates (green nodes in Figure~\ref{fig:top-k_illustration}) with the same prior condition, \ie only the prompt, which explores the \emph{breath} of the code generation. In contrast, the process of self-refinement is iteratively updating the prior condition, where the old generation and the execution results will be continuously updated and concatenated to the prompt as additional references (yellow nodes in Figure~\ref{fig:top-k_illustration}). Therefore, different from the breath exploration of top-k generation, self-refine is improving a specific generation with directional guidance in \emph{depth}.

We also conduct experiments to study how the orthogonal instincts perform differently during the code generation. 

\noindent\textbf{Setup.} We consider four types of top-k generation as a comparison: nucleus sampling with the temperature of (0.2, 0.6, 0.8), and beam search. Specifically, the code LM will generate the top-5 sequences as the prediction, where 5 is also the number of code \tool produces along the iterative self-refinement according to our configuration (Section~\ref{subsec:config_param}). Then we indexed the generated code for each method following their generation order. Finally, we check how many and which test cases each generation will pass individually across all methods. The experiment is conducted with the MBPP-S benchmark which contains 1,323 test cases in total.

\noindent\textbf{Findings.} As shown in Figure~\ref{fig:top_k_test_case_pass}, \tool eventually passes more test cases in MBPP-S step by step, since it learns to improve the old generation according to the execution results, revealing the behavior of exploring code generation in depth. On the contrary, the number of passed test cases is mostly similar across different generations of a specific top-k generation method, either nucleus sampling or beam search. After checking in detail, we realize that the accumulated number of passed test cases across all five generations is significantly higher than any individual generation. For example, all five generations from nucleus sampling with temperature 0.8 passed 513 test cases altogether, while one generation could only pass roughly 260 test cases. This reveals that top-k generation samples diverse code as the prediction, exploring the breadth of the search space.

\subsubsection{\tool could further refine top-k generations} 
\label{subsubsec:refine_top_k_results}
Motivated by the findings in \ref{subsubsec:top_k_is_orthogonal}, we delve deeper to study whether \tool's self-refinement is applicable to top-k generation and whether combining these two orthogonal techniques could further push the performance boundary. Specifically, \tool is supposed to be able to improve any past generation, so we evaluate \tool's performance to refine the top-k generation. As we show in Figure~\ref{fig:top_k_and_self_refine}, the top-k generations (green nodes) will be improved by \tool independently, producing refined programs for each of them (gray and yellow nodes).

\noindent\textbf{Setup.} We evaluate \tool's capacity in refining top-k generation using HumanEval, MBPP-S, and APPS benchmarks, similar to our main evaluation (Section~\ref{subsec: main_result}). We again choose nucleus sampling with the temperature of (0.2, 0.6, 0.8) and beam search to generate the top-5 predictions, and each generation will be refined four times, resulting in a total of twenty-five predictions.

\begin{table*}[!ht]
\centering
\caption{The test suite Pass@5 rate for three programming benchmarks. \tool's self-refinement is applicable to top-k generation and further improves its performance.}
\label{tab:self_refine_topk}
\resizebox{1.0\textwidth}{!}{
\begin{tabular}{l c c c c c c}
\toprule

\multirow{3}{*}{\textbf{Method}}  &\multicolumn{3}{c}{\textbf{One-time (Pass@5)}}&\multicolumn{3}{c}{\textbf{Self-Refine (Pass@5)}}\\\cmidrule(lr){2-4} \cmidrule(lr){5-7} 

&\textbf{ HumanEval} &\textbf{ MBPP-S} & \textbf{APPS} & \textbf{HumanEval} & \textbf{MBPP-S} & \textbf{APPS}\\\hline\hline

Beam Search & \colorgray 13.4 & \colorgray 21.8 & \colorgray 8.5 &\coloryellow 25.6~\improve{+91.0\%} & \coloryellow 47.3~\improve{+117.2\%} & \coloryellow 12.3~\improve{+44.0\%}\\\hdashline

Nuc. Samp. (tmp=0.2) & \colorgray 15.2 & \colorgray 26.7 & \colorgray 8.8 &\coloryellow 25.6~~\improve{+68.0\%} & \coloryellow 45.9~~\improve{+71.9\%} & \coloryellow 11.6~~\improve{+30.9\%} \\\hdashline

Nuc. Samp. (tmp=0.6) & \colorgray 18.9 & \colorgray 32.1 & \colorgray 9.1 &\coloryellow 28.1~\improve{+48.4\%} & \coloryellow 46.6~\improve{+45.3\%} & \coloryellow 11.9~\improve{+30.0\%}\\\hdashline%

Nuc. Samp. (tmp=0.8) & \colorgray 18.9 & \colorgray 29.7 & \colorgray 8.9 &\coloryellow 25.6~\improve{+35.4\%} & \coloryellow 47.3~\improve{+59.1\%} & \coloryellow 12.1~\improve{+36.1\%} \\

\hline\hline
\end{tabular}}
\end{table*} 

\noindent\textbf{Findings.} The results are reported in Table~\ref{tab:self_refine_topk}. We can see that \tool's self-refinement consistently and significantly boosts the pass@5 rate across four methods of top-k generation, up to 117.2\% relative improvement. This verifies that the self-refinement is compatible with the top-k generation and can further improve its performance.

\subsubsection{The overhead of \tool's self-refinement is comparable to top-k generation} 
\label{subsubsec:overhead_study}
Another difference between \tool's self-refinement and top-k generation is that, during the inference, the former is an iterative process while the latter is a simultaneous process. It is difficult to conceptually reason about which approach is supposed to have higher inference overhead since self-refinement is inevitably a sequential process, while top-k generation requires more GPU memory and additional computation of gathering the most probable tokens at each generation step.

\noindent\textbf{Setup.} To study and compare the overhead between \tool's self-refinement and top-k generation, we record two things during their inference: (1) the inference time (in seconds), and (2) the number of executions the machine performs to verify the correctness or collect the execution feedback. 

Note that the comparison is conducted based on the restriction that ``only five sequences can be generated", no matter whether the process is iterative or simultaneous. Consequently, top-k generation methods will produce five sequences with \emph{one-time inference}, while \tool produces five sequences by \emph{iteratively generating only one sequence} at a time, executing test cases and collecting feedback, and running inference again to refine the past generation.

\begin{table*}[!ht]
\centering
\caption{The comparison of inference overhead between \tool's self-refinement and top-k generation. ``Infer.(s)" represents the model inference time in seconds, ``\#Exec" represents the number of executions performed for evaluation, and ``Pass(\%)" represents the ratio of passed test suites (\ie the ratio of programming problems solved by the method)}
\label{tab:overhead_comparsion}
\resizebox{1\textwidth}{!}{
\begin{tabular}{l c c c c c c c c c}
\toprule

\multirow{3}{*}{\textbf{Method}}  &\multicolumn{3}{c}{\textbf{HumanEval}}&\multicolumn{3}{c}{\textbf{MBPP-S}} &\multicolumn{3}{c}{\textbf{APPS}}\\\cmidrule(lr){2-4} \cmidrule(lr){5-7} \cmidrule(lr){8-10} 

&\textbf{ Infer. (s)} &\textbf{ \#Exec.} & \textbf{Pass (\%)} & \textbf{ Infer. (s)} &\textbf{ \#Exec.} & \textbf{Pass (\%)} &\textbf{ Infer. (s)} &\textbf{ \#Exec.} & \textbf{Pass (\%)} \\\hline\hline

Beam Search@5 & 507.6 & 3263 & 13.4 & 1222.2 & 3473 & 21.8 & 4804.4 & 32472 & 8.5\\\hdashline%

Nuc. Samp.@5 (tmp=0.2) &425.5 & 4723 & 15.2 & 1084.4 & 5630 & 26.7 &5179.9 & 31189 & 8.8\\\hdashline

Nuc. Samp.@5 (tmp=0.6) &428.9 & 6422 & 18.9 & 1147.8 & 8224 & 32.1 &5213.7 & 32412 & 9.1\\\hdashline

Nuc. Samp.@5 (tmp=0.8) &423.9 & 6612 & 18.9 & 1084.7 & 8568 & 29.7 &5224.8 & 32458 & 8.9\\\hline\hline

\tool@5 &403.2 & 6725 & 20.7 &762.6 & 8755 & 32.3 &4238.3 & 32475 & 8.7\\

\hline\hline

\end{tabular}}
\end{table*}

\noindent\textbf{Findings.} We show the comparison in Table~\ref{tab:overhead_comparsion}. When we compare the inference time of different methods, we notice that \tool's inference time with self-refinement of generating five sequences is actually shorter than top-k generation, even if self-refinement is an iterative process. The main reason is that, restricted by the fixed size of GPU memory, the batch size of top-k generation has to be reduced when generating more sequences simultaneously, while self-refinement only generates one prediction at a time, so the batch size is significantly larger. The efficient parallel computation of GPU for a larger batch compensates for the time consumption of \tool's sequential refinement, while the smaller batch size of top-k generation results in more batches that have to be sequentially fed into GPU instead.

Conceptually, the number of executions of test cases should be exactly the same between top-k generation and \tool's self-refinement, since they generate the same number of predictions for each programming problem in the benchmarks, and the number of test cases for each problem is fixed. However, in practice, we notice that the numbers are varied across methods, and \tool is more comparable to nucleus sampling with a high temperature but higher than that with a low temperature and the beam search. The main reason is that our implementation of the execution framework follows the original design of HumanEval, where \citet{chen2021codex} minimizes the number of executions by executing the same predictions only once. As a result, the more diverse the predictions are, the more executions will be performed, and that is why we see an increase in the execution number when the temperature increases.

To conclude, we could not see a significant gap between top-k generation and \tool's self-refinement, in terms of both performance and inference overhead, since they have advantages and disadvantages in orthogonal angles. As we found in Section~\ref{subsubsec:refine_top_k_results}, the better solution is to take advantage of both techniques.

\begin{finding}
\tool's self-refinement is an orthogonal direction to top-k generation with comparable overhead, where the former improves one single generation in depth iteratively while the latter encourages exploring the breadth of the sequence space for the one-time generation. Further, \tool could be applied to top-k generation to further improve the overall performance of code generation.
\end{finding}

\subsection{RQ4. Ablation Study of Past Generation Mask and Data Mixture}
\label{subsec:ablation}

In this section, we conduct an ablation study to analyze the impact of Past Generation Mask's (PGM) masking ratio and the data combination for self-refinement training, since these empirical choices are expected to have non-trivial impacts on the learning process. 

As we introduced in Section~\ref{subsubsec:learning_to_self_refine}, PGM randomly masks a certain ratio of code tokens to prevent the model from lazily taking shortcuts, using the exact copy of the past, faulty generation as the refined prediction. However, different masking ratios could result in different learning behaviors. Concretely, when the masking ratio is high, there will be limited tokens for the model to refer to, and the model has to produce most tokens from scratch. On the contrary, when the masking ratio is low, the model could learn to borrow some helpful pieces of code, such as the code structure and the meaningful identifiers while not naively replicating past errors. 

In Section~\ref{subsubsec:mix_of_data}, we proposed to mix self-refine samples (Section~\ref{subsec:data_preparation}) and original code samples to construct the data for training, ensuring the model effectively learns self-refining while maintaining the capacity of general code generation. However, the proportion of these two data resources could affect the learning process. Specifically, when self-refine samples are overly dominant, the model will be optimized mostly towards self-refinement, where the natural language prompt, faulty generation, and the execution results are all required to present as the prior condition, while losing the capacity in one-time code generation where only the prompt is given, and vice versa.

\noindent\textbf{Setup.} To conduct the ablation study on the PGM masking ratio and the data combination proportion, we train \tool-350M multiple times with controlled settings and evaluate across the three programming benchmarks to compare across different settings for concluding the trend. Specifically, to study the impacts of PGM masking ratio we train \tool-350M four times with ratios of [0.0, 0.05, 0.15, 0.30] while maintaining all other training settings the same. Similarly, to study the difference across varied data mixture proportions, we train \tool-350M five times with the self-refine data ratios of [0\%, 25\%, 50\%, 75\%, 100\%].

\begin{figure}[!h]
  \centering
  \begin{subfigure}[b]{0.4\textwidth}
    \includegraphics[width=\textwidth]{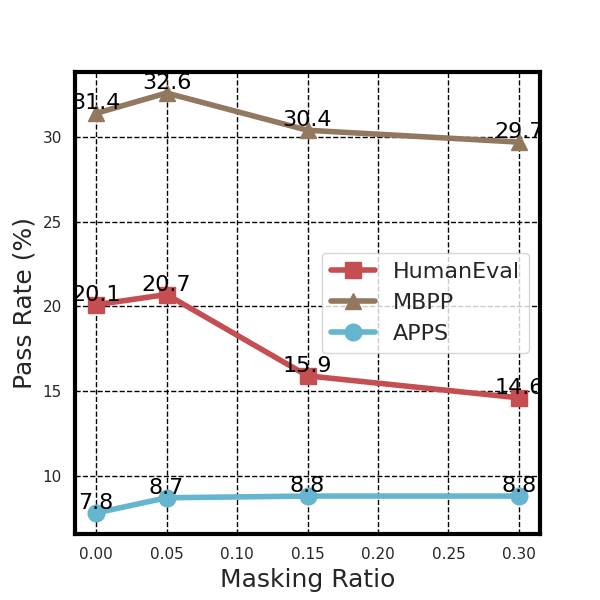}
    \caption{The impact of PGM masking ratio}
    \label{fig:pgm_ratio}
  \end{subfigure}
  \begin{subfigure}[b]{0.4\textwidth}
    \includegraphics[width=\textwidth]{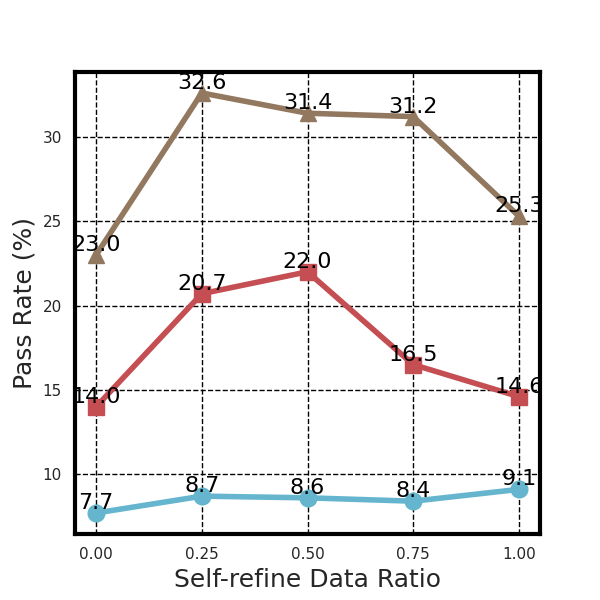}
    \caption{The impact of data mixture proportion}
    \label{fig:data_combine_ratio}
  \end{subfigure}
  \caption{Ablation Study}
  \label{fig:ablation}
  \vspace{4mm}
\end{figure}

\noindent\textbf{Findings.} The results regarding PGM masking ratios are plotted in Figure~\ref{fig:pgm_ratio}. From the figure, we deduce that PGM operates optimally at a lower masking ratio. Specifically, a masking ratio of around 0.05 emerges as an optimal point, indicating that the slight obfuscation can effectively prevent the exact copy during self-refinement while retaining valuable referring tokens in the faulty generation. However, going too high with the masking ratio seems to counteract its benefits, especially evident in the MBPP's sharp decline.

The results of the data mixture proportion are plotted in Figure~\ref{fig:data_combine_ratio}. We can see that an amalgamation of original and self-refined code samples offers superior training outcomes than one resource standalone. A balance, possibly around the 25\% ratio, appears to be a sweet spot. Solely depending on one kind of data, especially pure self-refined data, does not yield the best results, highlighting the importance of diverse training samples. Interestingly, we notice the performance on APPS exhibits a relatively stable pass rate, starting at 7.7\%, slightly peaking at 8.7\% for a 25\% data ratio, and settling around 9.1\% for a full self-refinement.
We speculate that the slightly higher performance in APPS when trained with 100\% self-refine samples is due to the similar distribution between our training data resources (Section~\ref{subsec:data_preparation}) and APPS, where both mostly contain more challenging programming problems than HumanEval and MBPP.

\begin{finding}
In conclusion, while PGM and data mixture are valuable techniques in the realm of self-refinement, careful calibration of their parameters, specifically masking ratio and data combination proportion, is essential for optimal performance.
\end{finding}

\section{Related Work}

\paragraph{\textbf{Pre-trained Code Large Language Models for Code Generation}}
The bloom of generative pre-trained large language models (LLMs) like GPT-3~\citep{brown2020language} in natural language processing has inspired a series of models that are pre-trained on source code with a focus on the task of code generation~\citep{zan-etal-2023-large}.
Codex~\citep{chen2021codex} is a series of proprietary Code LLMs up to 12B fine-tuned from GPT-3 on publicly available code from GitHub.
The 12B Codex-S, which is further fine-tuned on standalone functions achieves 37.7 Pass@1 on the HumanEval benchmark also proposed by~\citet{chen2021codex}. 
GitHub Copilot~\citep{github-2021-copilot}, a production version of Codex shows potential to assist software development through its code completion ability.

Following Codex, many open-source models have been proposed. 
CodeGen~\citep{nijkamp2023codegen} is a series of open-source Code LLMs up to 16B trained on English, multiple programming languages, and then Python datasets in order.
The 16B CodeGen achieves 29.28 Pass@1 on the HumanEval benchmark.
However, only left-to-right code completion doesn't cover all software engineering needs such as code debugging and refactoring.
Therefore, a class of models like Incoder~\citep{fried2023incoder} and SantaCoder~\citep{allal2023santacoder} explores another pre-training task, fill-in-the-middle (FIM), which tasks the model to fill in a masked region based on the surrounding context.
However, FIM models can only help with debugging or refactoring after human software engineers locate the regions of code with bugs or need improvement.
This doesn't align with real-life software development scenarios, where human engineers often spend the most time trying to pinpoint the regions that need to be refined.

Later, the success of ChatGPT~\citep{openai-2022-chatgpt}, a series of proprietary models that can follow instructions in a conversational way, opens up new opportunities to train Code LLMs to refine code in a flexible manner.
These models can be prompted to perform unlimited new open-ended generation tasks, even if they have not been trained on these tasks.
Particularly for code generation, users can prompt the chat models to edit, debug, and refactor existing code, especially code generated by themselves.
The technique behind those versatile chat models is
instruction tuning~\citep{mishra-etal-2022-cross, wei2022finetuned, sanh2022multitask}. It fine-tunes pre-trained LLMs on instruction-answer pairs covering a wide variety of tasks, which aims to generalize the instruction-following ability to unseen tasks.
Along this promising direction, most recent open-source models like StarCoder~\citep{li2023starcoder} and Code Llama~\citep{rozière2023codellama} are released with instruction-tuned versions in addition to pre-trained models.
Nonetheless, current instruction tuning procedures and datasets only focus on the one-step instruction-following ability but overlook the potential of raising the model's capability of refining answers in multiple steps.

\paragraph{\textbf{Improving the Quality of LLM Generation from Feedback}}
Instruction tuning not only empowers LLMs to multitask but also aligns their generation with human preference since the curated instruction-answer pairs in the datasets imply how humans would appropriately solve certain problems. 
However, instruction tuning is costly as it requires a large amount of high-quality human-labeled data.

To address this issue, a developed technique, Reinforcement Learning from Human Feedback (RLHF)~\citep{christiano2023deep,ziegler2020finetuning, stiennon2022learning, ouyang2022instructgpt} is adopted to further improve instruction-tuned LLMs.
Specifically, during the RLHF process, human annotators first grade the quality of various LLM generations by assigning a score, then a reward model is trained to emulate human grading, and finally, the LLM is fine-tuned using reinforcement learning with the reward model.
In this way, RLHF helps LLMs capture implicit human feedback information by training them for higher rewards. 
The most powerful chat models like GPT-4~\citep{openai2023gpt4}, Claude~\citep{anthropic-2023-claude}, and Llama2-Chat~\citep{touvron2023llama} are all fine-tuend using RLHF.
However, in the RLHF framework, the original human feedback is typically a single discrete score, which is implicit and sometimes even conflicting.
Moreover, improving generation using the RLHF requires updating the model's parameters, which cannot help enhance real-time user experience.
For example, within the RLHF framework, human software engineers can give a score to the model-generated code as real-time feedback, but it's not practical to update the model's parameters immediately and this single score doesn't provide any explicit guidance about how to improve future generations.

To complement the limitations of RLHF, efforts have been made to provide LLMs with explicit feedback as prompts and leverage their instruction-following ability to refine previously generated outputs~\citep{madaan2023selfrefine, shinn2023reflexion, nair2023dera}.
Typically, the explicit feedback is a critique or an explanation generated by the same model.
Therefore, these methods can be categorized as ``Self-Refinement with Prompting".
For code generation, in particular,~\citet{zhang2023selfedit} uses unit test outputs and error messages as explicit feedback for self-refinement.~\citet{chen2023teaching} further includes self-generated explanations and execution traces to the feedback.
Self-Refinement with Prompting does improve the quality of code generation, but it demands the model to have a strong instruction-following capability as well as a long context window to include different types of feedback.

\vspace{4mm}
\section{Limitations and Threats to Validity}
\vspace{4mm}

To prove the concept of the above goal with the best efficiency, we choose to focus \tool on Python language as its first attempt. Python has gained immense popularity in recent years and stands as one of the most widely used programming languages. Its reputation for being beginner-friendly, versatile, and having a clean, readable syntax makes it a natural choice for initial trials. However, this makes our current version of \tool not applicable to other languages, such as C++ and Java, and it is challenging to unify the execution across multiple languages due to the complicated dependency. We regard the multi-lingual version of \tool as an exciting extension of this work.

Another limitation of our framework is the dependency on canonical solutions. Grounding our framework in canonical solutions ensures closer proximity to correct code, but such solutions do not always exist. We have to rely on the coding challenge benchmark, such as CodeContest to collect such data, and such data is restricted by scope and might not be fully reflecting the complexity of the realistic development. 

In addition, we assume the existence of unit test cases for the generated code snippets. This assumption might not be generalizable to collecting large-scale data. While well-developed projects typically maintain such test cases, it could be difficult to automatically pair the function with its corresponding unit test due to the varied configurations across different projects.

In conclusion, while our self-refinement framework has made promising strides in enhancing code generation, there remain exciting opportunities for growth. Each limitation not only informs us of the challenges but also provides a clear path for future research and improvement. Our vision is to harness these insights to evolve and make our framework even more robust and adaptable to the dynamic world of coding.
\vspace{2mm}
\section{Conclusion}
\vspace{2mm}
In this paper, we propose \tool framework, making an attempt to teach code LMs to self-refine according to the faulty generation in the past and the execution feedback. We evaluate \tool's code generation capability with three popular programming benchmarks: HumanEval~\citep{chen2021codex}, MBPP-Sanitized~\citep{austin2021synthesis}, and APPS~\citep{hendrycks2021apps}. To illustrate the effectiveness and generalizability of \tool, we train four variants of \tool with varied parameter sizes ranging from 350M to 3B. From the evaluation results, we conclude that \tool is pretty effective at self-refinement, consistently boosting the code generation performance, by up to 63.5\% relative improvement, across four model sizes on all three benchmarks, while maintaining decent one-time generation capacity. With efficient self-refinement learning, \tool-350M outperforms StarCoder-1B across all three programming benchmarks, and \tool-1B matches the performance of StarCoder-3B. With the in-depth analysis, we also empirically reveal that \tool is effective at capturing execution feedback and has great potential to assist human developers with iterative programming.

\section*{Data-Availability Statement}
We release our code, data, and model checkpoints to encourage further exploration in this direction. The artifact that supports the results discussed in this paper is available at \url{https://github.com/ARiSE-Lab/CYCLE_OOPSLA_24}.
\section*{Acknowledgement}

We would like to thank the anonymous reviewers for their valuable feedback and comments. This work was supported in part by an IBM Ph.D. Fellowship, DARPA/NIWC-Pacific N66001-21-C4018, NSF CNS-2247370, CCF-2221943, CCF-2313055, CCF-1845893, and CCF-2107405. Any opinions, findings, conclusions, or recommendations expressed herein are those of the authors and do not necessarily reflect those of IBM, DARPA, or NSF.

\bibliography{main}


\begin{thebibliography}{57}


\ifx \showCODEN    \undefined \def \showCODEN     #1{\unskip}     \fi
\ifx \showDOI      \undefined \def \showDOI       #1{#1}\fi
\ifx \showISBNx    \undefined \def \showISBNx     #1{\unskip}     \fi
\ifx \showISBNxiii \undefined \def \showISBNxiii  #1{\unskip}     \fi
\ifx \showISSN     \undefined \def \showISSN      #1{\unskip}     \fi
\ifx \showLCCN     \undefined \def \showLCCN      #1{\unskip}     \fi
\ifx \shownote     \undefined \def \shownote      #1{#1}          \fi
\ifx \showarticletitle \undefined \def \showarticletitle #1{#1}   \fi
\ifx \showURL      \undefined \def \showURL       {\relax}        \fi
\providecommand\bibfield[2]{#2}
\providecommand\bibinfo[2]{#2}
\providecommand\natexlab[1]{#1}
\providecommand\showeprint[2][]{arXiv:#2}

\bibitem[Allal et~al\mbox{.}(2023)]%
        {allal2023santacoder}
\bibfield{author}{\bibinfo{person}{Loubna~Ben Allal}, \bibinfo{person}{Raymond Li}, \bibinfo{person}{Denis Kocetkov}, \bibinfo{person}{Chenghao Mou}, \bibinfo{person}{Christopher Akiki}, \bibinfo{person}{Carlos~Munoz Ferrandis}, \bibinfo{person}{Niklas Muennighoff}, \bibinfo{person}{Mayank Mishra}, \bibinfo{person}{Alex Gu}, \bibinfo{person}{Manan Dey}, \bibinfo{person}{Logesh~Kumar Umapathi}, \bibinfo{person}{Carolyn~Jane Anderson}, \bibinfo{person}{Yangtian Zi}, \bibinfo{person}{Joel~Lamy Poirier}, \bibinfo{person}{Hailey Schoelkopf}, \bibinfo{person}{Sergey Troshin}, \bibinfo{person}{Dmitry Abulkhanov}, \bibinfo{person}{Manuel Romero}, \bibinfo{person}{Michael Lappert}, \bibinfo{person}{Francesco~De Toni}, \bibinfo{person}{Bernardo~García del Río}, \bibinfo{person}{Qian Liu}, \bibinfo{person}{Shamik Bose}, \bibinfo{person}{Urvashi Bhattacharyya}, \bibinfo{person}{Terry~Yue Zhuo}, \bibinfo{person}{Ian Yu}, \bibinfo{person}{Paulo Villegas}, \bibinfo{person}{Marco Zocca}, \bibinfo{person}{Sourab
  Mangrulkar}, \bibinfo{person}{David Lansky}, \bibinfo{person}{Huu Nguyen}, \bibinfo{person}{Danish Contractor}, \bibinfo{person}{Luis Villa}, \bibinfo{person}{Jia Li}, \bibinfo{person}{Dzmitry Bahdanau}, \bibinfo{person}{Yacine Jernite}, \bibinfo{person}{Sean Hughes}, \bibinfo{person}{Daniel Fried}, \bibinfo{person}{Arjun Guha}, \bibinfo{person}{Harm de Vries}, {and} \bibinfo{person}{Leandro von Werra}.} \bibinfo{year}{2023}\natexlab{}.
\newblock \bibinfo{title}{SantaCoder: don't reach for the stars!}
\newblock
\newblock
\showeprint[arxiv]{2301.03988}~[cs.SE]


\bibitem[Amazon(2023)]%
        {amazon-2023-codewhisperer}
\bibfield{author}{\bibinfo{person}{Amazon}.} \bibinfo{year}{2023}\natexlab{}.
\newblock \bibinfo{title}{Amazon CodeWhisperer: Build applications faster and more securely with your AI coding companion}.
\newblock \bibinfo{howpublished}{\url{https://aws.amazon.com/codewhisperer/}}.
\newblock


\bibitem[Anthropic(2023)]%
        {anthropic-2023-claude}
\bibfield{author}{\bibinfo{person}{Anthropic}.} \bibinfo{year}{2023}\natexlab{}.
\newblock \bibinfo{title}{Introducing Claude}.
\newblock \bibinfo{howpublished}{\url{https://www.anthropic.com/index/introducing-claude}}.
\newblock


\bibitem[Austin et~al\mbox{.}(2021)]%
        {austin2021synthesis}
\bibfield{author}{\bibinfo{person}{Jacob Austin}, \bibinfo{person}{Augustus Odena}, \bibinfo{person}{Maxwell Nye}, \bibinfo{person}{Maarten Bosma}, \bibinfo{person}{Henryk Michalewski}, \bibinfo{person}{David Dohan}, \bibinfo{person}{Ellen Jiang}, \bibinfo{person}{Carrie~J. Cai}, \bibinfo{person}{Michael Terry}, \bibinfo{person}{Quoc~V. Le}, {and} \bibinfo{person}{Charles Sutton}.} \bibinfo{year}{2021}\natexlab{}.
\newblock \showarticletitle{Program Synthesis with Large Language Models}.
\newblock \bibinfo{journal}{\emph{CoRR}}  \bibinfo{volume}{abs/2108.07732} (\bibinfo{year}{2021}).
\newblock
\showeprint[arXiv]{2108.07732}
\urldef\tempurl%
\url{https://arxiv.org/abs/2108.07732}
\showURL{%
\tempurl}


\bibitem[Barke et~al\mbox{.}(2023)]%
        {barke2023grounded}
\bibfield{author}{\bibinfo{person}{Shraddha Barke}, \bibinfo{person}{Michael~B. James}, {and} \bibinfo{person}{Nadia Polikarpova}.} \bibinfo{year}{2023}\natexlab{}.
\newblock \showarticletitle{Grounded Copilot: How Programmers Interact with Code-Generating Models}.
\newblock  \bibinfo{volume}{7}, \bibinfo{number}{OOPSLA1}, Article \bibinfo{articleno}{78} (\bibinfo{year}{2023}), \bibinfo{numpages}{27}~pages.
\newblock
\urldef\tempurl%
\url{https://doi.org/10.1145/3586030}
\showDOI{\tempurl}


\bibitem[Bavarian et~al\mbox{.}(2022)]%
        {bavarian2022efficient}
\bibfield{author}{\bibinfo{person}{Mohammad Bavarian}, \bibinfo{person}{Heewoo Jun}, \bibinfo{person}{Nikolas Tezak}, \bibinfo{person}{John Schulman}, \bibinfo{person}{Christine McLeavey}, \bibinfo{person}{Jerry Tworek}, {and} \bibinfo{person}{Mark Chen}.} \bibinfo{year}{2022}\natexlab{}.
\newblock \bibinfo{title}{Efficient Training of Language Models to Fill in the Middle}.
\newblock
\newblock
\showeprint[arxiv]{2207.14255}~[cs.CL]


\bibitem[Brown et~al\mbox{.}(2020)]%
        {brown2020language}
\bibfield{author}{\bibinfo{person}{Tom~B. Brown}, \bibinfo{person}{Benjamin Mann}, \bibinfo{person}{Nick Ryder}, \bibinfo{person}{Melanie Subbiah}, \bibinfo{person}{Jared Kaplan}, \bibinfo{person}{Prafulla Dhariwal}, \bibinfo{person}{Arvind Neelakantan}, \bibinfo{person}{Pranav Shyam}, \bibinfo{person}{Girish Sastry}, \bibinfo{person}{Amanda Askell}, \bibinfo{person}{Sandhini Agarwal}, \bibinfo{person}{Ariel Herbert-Voss}, \bibinfo{person}{Gretchen Krueger}, \bibinfo{person}{Tom Henighan}, \bibinfo{person}{Rewon Child}, \bibinfo{person}{Aditya Ramesh}, \bibinfo{person}{Daniel~M. Ziegler}, \bibinfo{person}{Jeffrey Wu}, \bibinfo{person}{Clemens Winter}, \bibinfo{person}{Christopher Hesse}, \bibinfo{person}{Mark Chen}, \bibinfo{person}{Eric Sigler}, \bibinfo{person}{Mateusz Litwin}, \bibinfo{person}{Scott Gray}, \bibinfo{person}{Benjamin Chess}, \bibinfo{person}{Jack Clark}, \bibinfo{person}{Christopher Berner}, \bibinfo{person}{Sam McCandlish}, \bibinfo{person}{Alec Radford}, \bibinfo{person}{Ilya Sutskever},
  {and} \bibinfo{person}{Dario Amodei}.} \bibinfo{year}{2020}\natexlab{}.
\newblock \bibinfo{title}{Language Models are Few-Shot Learners}.
\newblock
\newblock
\showeprint[arxiv]{2005.14165}~[cs.CL]


\bibitem[Chen et~al\mbox{.}(2021)]%
        {chen2021codex}
\bibfield{author}{\bibinfo{person}{Mark Chen}, \bibinfo{person}{Jerry Tworek}, \bibinfo{person}{Heewoo Jun}, \bibinfo{person}{Qiming Yuan}, \bibinfo{person}{Henrique~Pond{\'{e}} de Oliveira~Pinto}, {and} \bibinfo{person}{\etal}.} \bibinfo{year}{2021}\natexlab{}.
\newblock \showarticletitle{Evaluating Large Language Models Trained on Code}.
\newblock \bibinfo{journal}{\emph{CoRR}}  \bibinfo{volume}{abs/2107.03374} (\bibinfo{year}{2021}).
\newblock
\showeprint[arXiv]{2107.03374}
\urldef\tempurl%
\url{https://arxiv.org/abs/2107.03374}
\showURL{%
\tempurl}


\bibitem[Chen et~al\mbox{.}(2023)]%
        {chen2023teaching}
\bibfield{author}{\bibinfo{person}{Xinyun Chen}, \bibinfo{person}{Maxwell Lin}, \bibinfo{person}{Nathanael Schärli}, {and} \bibinfo{person}{Denny Zhou}.} \bibinfo{year}{2023}\natexlab{}.
\newblock \bibinfo{title}{Teaching Large Language Models to Self-Debug}.
\newblock
\newblock
\showeprint[arxiv]{2304.05128}~[cs.CL]


\bibitem[Christiano et~al\mbox{.}(2023)]%
        {christiano2023deep}
\bibfield{author}{\bibinfo{person}{Paul Christiano}, \bibinfo{person}{Jan Leike}, \bibinfo{person}{Tom~B. Brown}, \bibinfo{person}{Miljan Martic}, \bibinfo{person}{Shane Legg}, {and} \bibinfo{person}{Dario Amodei}.} \bibinfo{year}{2023}\natexlab{}.
\newblock \bibinfo{title}{Deep reinforcement learning from human preferences}.
\newblock
\newblock
\showeprint[arxiv]{1706.03741}~[stat.ML]


\bibitem[Ding et~al\mbox{.}(2020)]%
        {ding2020patching}
\bibfield{author}{\bibinfo{person}{Yangruibo Ding}, \bibinfo{person}{Baishakhi Ray}, \bibinfo{person}{Devanbu Premkumar}, {and} \bibinfo{person}{Vincent~J. Hellendoorn}.} \bibinfo{year}{2020}\natexlab{}.
\newblock \showarticletitle{Patching as Translation: the Data and the Metaphor}. In \bibinfo{booktitle}{\emph{35th IEEE/ACM International Conference on Automated Software Engineering}} (Virtual Event, Australia) \emph{(\bibinfo{series}{ASE '20})}.
\newblock
\urldef\tempurl%
\url{https://doi.org/10.1145/3324884.3416587}
\showDOI{\tempurl}


\bibitem[\etal(2019)]%
        {paszke2019pytorch}
\bibfield{author}{\bibinfo{person}{Adam~Paszke \etal}.} \bibinfo{year}{2019}\natexlab{}.
\newblock \bibinfo{booktitle}{\emph{PyTorch: An Imperative Style, High-Performance Deep Learning Library}}.
\newblock


\bibitem[\etal(2020)]%
        {wolf-etal-2020-huggingface}
\bibfield{author}{\bibinfo{person}{Thomas~Wolf \etal}.} \bibinfo{year}{2020}\natexlab{}.
\newblock \showarticletitle{Transformers: State-of-the-Art Natural Language Processing}. In \bibinfo{booktitle}{\emph{Proceedings of the 2020 Conference on Empirical Methods in Natural Language Processing: System Demonstrations}}. \bibinfo{publisher}{Association for Computational Linguistics}, \bibinfo{address}{Online}, \bibinfo{pages}{38--45}.
\newblock
\urldef\tempurl%
\url{https://doi.org/10.18653/v1/2020.emnlp-demos.6}
\showDOI{\tempurl}


\bibitem[Fried et~al\mbox{.}(2023)]%
        {fried2023incoder}
\bibfield{author}{\bibinfo{person}{Daniel Fried}, \bibinfo{person}{Armen Aghajanyan}, \bibinfo{person}{Jessy Lin}, \bibinfo{person}{Sida Wang}, \bibinfo{person}{Eric Wallace}, \bibinfo{person}{Freda Shi}, \bibinfo{person}{Ruiqi Zhong}, \bibinfo{person}{Wen tau Yih}, \bibinfo{person}{Luke Zettlemoyer}, {and} \bibinfo{person}{Mike Lewis}.} \bibinfo{year}{2023}\natexlab{}.
\newblock \bibinfo{title}{InCoder: A Generative Model for Code Infilling and Synthesis}.
\newblock
\newblock
\showeprint[arxiv]{2204.05999}~[cs.SE]


\bibitem[Gao et~al\mbox{.}(2021)]%
        {gao2021thepile}
\bibfield{author}{\bibinfo{person}{Leo Gao}, \bibinfo{person}{Stella Biderman}, \bibinfo{person}{Sid Black}, \bibinfo{person}{Laurence Golding}, \bibinfo{person}{Travis Hoppe}, \bibinfo{person}{Charles Foster}, \bibinfo{person}{Jason Phang}, \bibinfo{person}{Horace He}, \bibinfo{person}{Anish Thite}, \bibinfo{person}{Noa Nabeshima}, \bibinfo{person}{Shawn Presser}, {and} \bibinfo{person}{Connor Leahy}.} \bibinfo{year}{2021}\natexlab{}.
\newblock \showarticletitle{The Pile: An 800GB Dataset of Diverse Text for Language Modeling}.
\newblock \bibinfo{journal}{\emph{CoRR}}  \bibinfo{volume}{abs/2101.00027} (\bibinfo{year}{2021}).
\newblock
\showeprint[arXiv]{2101.00027}
\urldef\tempurl%
\url{https://arxiv.org/abs/2101.00027}
\showURL{%
\tempurl}


\bibitem[GitHub(2021)]%
        {github-2021-copilot}
\bibfield{author}{\bibinfo{person}{GitHub}.} \bibinfo{year}{2021}\natexlab{}.
\newblock \bibinfo{title}{GitHub Copilot: Your AI Pair Programmer}.
\newblock \bibinfo{howpublished}{\url{https://copilot.github.com/}}.
\newblock


\bibitem[Guo et~al\mbox{.}(2023)]%
        {guo2023exploring}
\bibfield{author}{\bibinfo{person}{Qi Guo}, \bibinfo{person}{Junming Cao}, \bibinfo{person}{Xiaofei Xie}, \bibinfo{person}{Shangqing Liu}, \bibinfo{person}{Xiaohong Li}, \bibinfo{person}{Bihuan Chen}, {and} \bibinfo{person}{Xin Peng}.} \bibinfo{year}{2023}\natexlab{}.
\newblock \bibinfo{title}{Exploring the Potential of ChatGPT in Automated Code Refinement: An Empirical Study}.
\newblock
\newblock
\showeprint[arxiv]{2309.08221}~[cs.SE]


\bibitem[He and Vechev(2023)]%
        {sven-llm}
\bibfield{author}{\bibinfo{person}{Jingxuan He} {and} \bibinfo{person}{Martin Vechev}.} \bibinfo{year}{2023}\natexlab{}.
\newblock \showarticletitle{Large Language Models for Code: Security Hardening and Adversarial Testing}. In \bibinfo{booktitle}{\emph{Proceedings of the 2023 ACM SIGSAC Conference on Computer and Communications Security (CCS’23)}}.
\newblock


\bibitem[Hendrycks et~al\mbox{.}(2021)]%
        {hendrycks2021apps}
\bibfield{author}{\bibinfo{person}{Dan Hendrycks}, \bibinfo{person}{Steven Basart}, \bibinfo{person}{Saurav Kadavath}, \bibinfo{person}{Mantas Mazeika}, \bibinfo{person}{Akul Arora}, \bibinfo{person}{Ethan Guo}, \bibinfo{person}{Collin Burns}, \bibinfo{person}{Samir Puranik}, \bibinfo{person}{Horace He}, \bibinfo{person}{Dawn Song}, {and} \bibinfo{person}{Jacob Steinhardt}.} \bibinfo{year}{2021}\natexlab{}.
\newblock \showarticletitle{Measuring Coding Challenge Competence With APPS}. In \bibinfo{booktitle}{\emph{Proceedings of the Neural Information Processing Systems Track on Datasets and Benchmarks}}, \bibfield{editor}{\bibinfo{person}{J.~Vanschoren} {and} \bibinfo{person}{S.~Yeung}} (Eds.), Vol.~\bibinfo{volume}{1}. \bibinfo{publisher}{Curran}.
\newblock
\urldef\tempurl%
\url{https://datasets-benchmarks-proceedings.neurips.cc/paper_files/paper/2021/file/c24cd76e1ce41366a4bbe8a49b02a028-Paper-round2.pdf}
\showURL{%
\tempurl}


\bibitem[Holtzman et~al\mbox{.}(2020)]%
        {Holtzman2020nucleussampling}
\bibfield{author}{\bibinfo{person}{Ari Holtzman}, \bibinfo{person}{Jan Buys}, \bibinfo{person}{Li Du}, \bibinfo{person}{Maxwell Forbes}, {and} \bibinfo{person}{Yejin Choi}.} \bibinfo{year}{2020}\natexlab{}.
\newblock \showarticletitle{The Curious Case of Neural Text Degeneration}. In \bibinfo{booktitle}{\emph{International Conference on Learning Representations}}.
\newblock
\urldef\tempurl%
\url{https://openreview.net/forum?id=rygGQyrFvH}
\showURL{%
\tempurl}


\bibitem[Huang et~al\mbox{.}(2023)]%
        {huang2023large}
\bibfield{author}{\bibinfo{person}{Jie Huang}, \bibinfo{person}{Xinyun Chen}, \bibinfo{person}{Swaroop Mishra}, \bibinfo{person}{Huaixiu~Steven Zheng}, \bibinfo{person}{Adams~Wei Yu}, \bibinfo{person}{Xinying Song}, {and} \bibinfo{person}{Denny Zhou}.} \bibinfo{year}{2023}\natexlab{}.
\newblock \bibinfo{title}{Large Language Models Cannot Self-Correct Reasoning Yet}.
\newblock
\newblock
\showeprint[arxiv]{2310.01798}~[cs.CL]


\bibitem[HuggingFace(2023)]%
        {hfmodelhub}
\bibfield{author}{\bibinfo{person}{HuggingFace}.} \bibinfo{year}{2023}\natexlab{}.
\newblock \bibinfo{booktitle}{\emph{Hugging Face Model Hub}}.
\newblock
\newblock
\shownote{\url{https://huggingface.co/models}}.


\bibitem[Kaplan et~al\mbox{.}(2020)]%
        {kaplan2020scaling}
\bibfield{author}{\bibinfo{person}{Jared Kaplan}, \bibinfo{person}{Sam McCandlish}, \bibinfo{person}{Tom Henighan}, \bibinfo{person}{Tom~B. Brown}, \bibinfo{person}{Benjamin Chess}, \bibinfo{person}{Rewon Child}, \bibinfo{person}{Scott Gray}, \bibinfo{person}{Alec Radford}, \bibinfo{person}{Jeffrey Wu}, {and} \bibinfo{person}{Dario Amodei}.} \bibinfo{year}{2020}\natexlab{}.
\newblock \bibinfo{title}{Scaling Laws for Neural Language Models}.
\newblock
\newblock
\showeprint[arxiv]{2001.08361}~[cs.LG]


\bibitem[Kocetkov et~al\mbox{.}(2022)]%
        {kocetkov2022stack}
\bibfield{author}{\bibinfo{person}{Denis Kocetkov}, \bibinfo{person}{Raymond Li}, \bibinfo{person}{Loubna~Ben Allal}, \bibinfo{person}{Jia Li}, \bibinfo{person}{Chenghao Mou}, \bibinfo{person}{Carlos~Mu{\~n}oz Ferrandis}, \bibinfo{person}{Yacine Jernite}, \bibinfo{person}{Margaret Mitchell}, \bibinfo{person}{Sean Hughes}, \bibinfo{person}{Thomas Wolf}, {et~al\mbox{.}}} \bibinfo{year}{2022}\natexlab{}.
\newblock \showarticletitle{The Stack: 3 TB of permissively licensed source code}.
\newblock \bibinfo{journal}{\emph{arXiv preprint arXiv:2211.15533}} (\bibinfo{year}{2022}).
\newblock


\bibitem[Kudo and Richardson(2018)]%
        {kudo-richardson-2018-sentencepiece}
\bibfield{author}{\bibinfo{person}{Taku Kudo} {and} \bibinfo{person}{John Richardson}.} \bibinfo{year}{2018}\natexlab{}.
\newblock \showarticletitle{{S}entence{P}iece: A simple and language independent subword tokenizer and detokenizer for Neural Text Processing}. In \bibinfo{booktitle}{\emph{Proceedings of the 2018 Conference on Empirical Methods in Natural Language Processing: System Demonstrations}}. \bibinfo{publisher}{Association for Computational Linguistics}, \bibinfo{address}{Brussels, Belgium}, \bibinfo{pages}{66--71}.
\newblock
\urldef\tempurl%
\url{https://doi.org/10.18653/v1/D18-2012}
\showDOI{\tempurl}


\bibitem[Li et~al\mbox{.}(2023)]%
        {li2023starcoder}
\bibfield{author}{\bibinfo{person}{Raymond Li}, \bibinfo{person}{Loubna~Ben Allal}, \bibinfo{person}{Yangtian Zi}, \bibinfo{person}{Niklas Muennighoff}, \bibinfo{person}{Denis Kocetkov}, \bibinfo{person}{Chenghao Mou}, {and} \bibinfo{person}{\etal}.} \bibinfo{year}{2023}\natexlab{}.
\newblock \bibinfo{title}{StarCoder: may the source be with you!}
\newblock
\newblock
\showeprint[arxiv]{2305.06161}~[cs.CL]


\bibitem[Li et~al\mbox{.}(2022)]%
        {li2022alphacode}
\bibfield{author}{\bibinfo{person}{Yujia Li}, \bibinfo{person}{David Choi}, \bibinfo{person}{Junyoung Chung}, \bibinfo{person}{Nate Kushman}, \bibinfo{person}{Julian Schrittwieser}, \bibinfo{person}{Rémi Leblond}, \bibinfo{person}{Tom Eccles}, {and} \bibinfo{person}{\etal}.} \bibinfo{year}{2022}\natexlab{}.
\newblock \showarticletitle{Competition-level code generation with AlphaCode}.
\newblock \bibinfo{journal}{\emph{Science}} \bibinfo{volume}{378}, \bibinfo{number}{6624} (\bibinfo{year}{2022}), \bibinfo{pages}{1092--1097}.
\newblock
\urldef\tempurl%
\url{https://doi.org/10.1126/science.abq1158}
\showDOI{\tempurl}
\showeprint{https://www.science.org/doi/pdf/10.1126/science.abq1158}


\bibitem[Liu et~al\mbox{.}(2023)]%
        {liu2023forgetful}
\bibfield{author}{\bibinfo{person}{Hao Liu}, \bibinfo{person}{Xinyang Geng}, \bibinfo{person}{Lisa Lee}, \bibinfo{person}{Igor Mordatch}, \bibinfo{person}{Sergey Levine}, \bibinfo{person}{Sharan Narang}, {and} \bibinfo{person}{Pieter Abbeel}.} \bibinfo{year}{2023}\natexlab{}.
\newblock \bibinfo{title}{Forgetful causal masking makes causal language models better zero-shot learners}.
\newblock
\newblock
\urldef\tempurl%
\url{https://openreview.net/forum?id=YrZEKNLWhlp}
\showURL{%
\tempurl}


\bibitem[Madaan et~al\mbox{.}(2023)]%
        {madaan2023selfrefine}
\bibfield{author}{\bibinfo{person}{Aman Madaan}, \bibinfo{person}{Niket Tandon}, \bibinfo{person}{Prakhar Gupta}, \bibinfo{person}{Skyler Hallinan}, \bibinfo{person}{Luyu Gao}, \bibinfo{person}{Sarah Wiegreffe}, \bibinfo{person}{Uri Alon}, \bibinfo{person}{Nouha Dziri}, \bibinfo{person}{Shrimai Prabhumoye}, \bibinfo{person}{Yiming Yang}, \bibinfo{person}{Shashank Gupta}, \bibinfo{person}{Bodhisattwa~Prasad Majumder}, \bibinfo{person}{Katherine Hermann}, \bibinfo{person}{Sean Welleck}, \bibinfo{person}{Amir Yazdanbakhsh}, {and} \bibinfo{person}{Peter Clark}.} \bibinfo{year}{2023}\natexlab{}.
\newblock \bibinfo{title}{Self-Refine: Iterative Refinement with Self-Feedback}.
\newblock
\newblock
\showeprint[arxiv]{2303.17651}~[cs.CL]


\bibitem[Mirzayanov(2020)]%
        {codeforces2020}
\bibfield{author}{\bibinfo{person}{Mike Mirzayanov}.} \bibinfo{year}{2020}\natexlab{}.
\newblock \bibinfo{booktitle}{\emph{Codeforces: Results of 2020}}.
\newblock
\newblock
\shownote{\url{https://codeforces.com/blog/entry/89502}}.


\bibitem[Mishra et~al\mbox{.}(2022)]%
        {mishra-etal-2022-cross}
\bibfield{author}{\bibinfo{person}{Swaroop Mishra}, \bibinfo{person}{Daniel Khashabi}, \bibinfo{person}{Chitta Baral}, {and} \bibinfo{person}{Hannaneh Hajishirzi}.} \bibinfo{year}{2022}\natexlab{}.
\newblock \showarticletitle{Cross-Task Generalization via Natural Language Crowdsourcing Instructions}. In \bibinfo{booktitle}{\emph{Proceedings of the 60th Annual Meeting of the Association for Computational Linguistics (Volume 1: Long Papers)}}. \bibinfo{publisher}{Association for Computational Linguistics}, \bibinfo{address}{Dublin, Ireland}, \bibinfo{pages}{3470--3487}.
\newblock
\urldef\tempurl%
\url{https://doi.org/10.18653/v1/2022.acl-long.244}
\showDOI{\tempurl}


\bibitem[Nair et~al\mbox{.}(2023)]%
        {nair2023dera}
\bibfield{author}{\bibinfo{person}{Varun Nair}, \bibinfo{person}{Elliot Schumacher}, \bibinfo{person}{Geoffrey Tso}, {and} \bibinfo{person}{Anitha Kannan}.} \bibinfo{year}{2023}\natexlab{}.
\newblock \bibinfo{title}{DERA: Enhancing Large Language Model Completions with Dialog-Enabled Resolving Agents}.
\newblock
\newblock
\showeprint[arxiv]{2303.17071}~[cs.CL]


\bibitem[Nijkamp et~al\mbox{.}(2023a)]%
        {nijkamp2023codegen2}
\bibfield{author}{\bibinfo{person}{Erik Nijkamp}, \bibinfo{person}{Hiroaki Hayashi}, \bibinfo{person}{Caiming Xiong}, \bibinfo{person}{Silvio Savarese}, {and} \bibinfo{person}{Yingbo Zhou}.} \bibinfo{year}{2023}\natexlab{a}.
\newblock \bibinfo{title}{CodeGen2: Lessons for Training LLMs on Programming and Natural Languages}.
\newblock
\newblock
\showeprint[arxiv]{2305.02309}~[cs.LG]


\bibitem[Nijkamp et~al\mbox{.}(2023b)]%
        {nijkamp2023codegen}
\bibfield{author}{\bibinfo{person}{Erik Nijkamp}, \bibinfo{person}{Bo Pang}, \bibinfo{person}{Hiroaki Hayashi}, \bibinfo{person}{Lifu Tu}, \bibinfo{person}{Huan Wang}, \bibinfo{person}{Yingbo Zhou}, \bibinfo{person}{Silvio Savarese}, {and} \bibinfo{person}{Caiming Xiong}.} \bibinfo{year}{2023}\natexlab{b}.
\newblock \showarticletitle{CodeGen: An Open Large Language Model for Code with Multi-Turn Program Synthesis}. In \bibinfo{booktitle}{\emph{The Eleventh International Conference on Learning Representations}}.
\newblock
\urldef\tempurl%
\url{https://openreview.net/forum?id=iaYcJKpY2B_}
\showURL{%
\tempurl}


\bibitem[OpenAI(2022)]%
        {openai-2022-chatgpt}
\bibfield{author}{\bibinfo{person}{OpenAI}.} \bibinfo{year}{2022}\natexlab{}.
\newblock \bibinfo{title}{Introducing ChatGPT}.
\newblock \bibinfo{howpublished}{\url{https://openai.com/blog/chatgpt/}}.
\newblock


\bibitem[OpenAI(2023)]%
        {openai2023gpt4}
\bibfield{author}{\bibinfo{person}{OpenAI}.} \bibinfo{year}{2023}\natexlab{}.
\newblock \bibinfo{title}{GPT-4 Technical Report}.
\newblock
\newblock
\showeprint[arxiv]{2303.08774}~[cs.CL]


\bibitem[Ouyang et~al\mbox{.}(2022)]%
        {ouyang2022instructgpt}
\bibfield{author}{\bibinfo{person}{Long Ouyang}, \bibinfo{person}{Jeff Wu}, \bibinfo{person}{Xu Jiang}, \bibinfo{person}{Diogo Almeida}, \bibinfo{person}{Carroll~L. Wainwright}, \bibinfo{person}{Pamela Mishkin}, \bibinfo{person}{Chong Zhang}, \bibinfo{person}{Sandhini Agarwal}, \bibinfo{person}{Katarina Slama}, \bibinfo{person}{Alex Ray}, \bibinfo{person}{John Schulman}, \bibinfo{person}{Jacob Hilton}, \bibinfo{person}{Fraser Kelton}, \bibinfo{person}{Luke Miller}, \bibinfo{person}{Maddie Simens}, \bibinfo{person}{Amanda Askell}, \bibinfo{person}{Peter Welinder}, \bibinfo{person}{Paul Christiano}, \bibinfo{person}{Jan Leike}, {and} \bibinfo{person}{Ryan Lowe}.} \bibinfo{year}{2022}\natexlab{}.
\newblock \bibinfo{title}{Training language models to follow instructions with human feedback}.
\newblock
\newblock
\showeprint[arxiv]{2203.02155}~[cs.CL]


\bibitem[Puri et~al\mbox{.}(2021)]%
        {puri2021codenet}
\bibfield{author}{\bibinfo{person}{Ruchir Puri}, \bibinfo{person}{David~S. Kung}, \bibinfo{person}{Geert Janssen}, \bibinfo{person}{Wei Zhang}, \bibinfo{person}{Giacomo Domeniconi}, \bibinfo{person}{Vladimir Zolotov}, \bibinfo{person}{Julian Dolby}, \bibinfo{person}{Jie Chen}, \bibinfo{person}{Mihir~R. Choudhury}, \bibinfo{person}{Lindsey Decker}, \bibinfo{person}{Veronika Thost}, \bibinfo{person}{Luca Buratti}, \bibinfo{person}{Saurabh Pujar}, {and} \bibinfo{person}{Ulrich Finkler}.} \bibinfo{year}{2021}\natexlab{}.
\newblock \showarticletitle{Project CodeNet: {A} Large-Scale {AI} for Code Dataset for Learning a Diversity of Coding Tasks}.
\newblock \bibinfo{journal}{\emph{CoRR}}  \bibinfo{volume}{abs/2105.12655} (\bibinfo{year}{2021}).
\newblock
\showeprint[arXiv]{2105.12655}
\urldef\tempurl%
\url{https://arxiv.org/abs/2105.12655}
\showURL{%
\tempurl}


\bibitem[Radford and Narasimhan(2018)]%
        {Radford2018gpt1}
\bibfield{author}{\bibinfo{person}{Alec Radford} {and} \bibinfo{person}{Karthik Narasimhan}.} \bibinfo{year}{2018}\natexlab{}.
\newblock \showarticletitle{Improving Language Understanding by Generative Pre-Training}.
\newblock
\urldef\tempurl%
\url{https://api.semanticscholar.org/CorpusID:49313245}
\showURL{%
\tempurl}


\bibitem[Radford et~al\mbox{.}(2019)]%
        {Radford2019gpt2}
\bibfield{author}{\bibinfo{person}{Alec Radford}, \bibinfo{person}{Jeff Wu}, \bibinfo{person}{Rewon Child}, \bibinfo{person}{David Luan}, \bibinfo{person}{Dario Amodei}, {and} \bibinfo{person}{Ilya Sutskever}.} \bibinfo{year}{2019}\natexlab{}.
\newblock \showarticletitle{Language Models are Unsupervised Multitask Learners}.
\newblock
\urldef\tempurl%
\url{https://api.semanticscholar.org/CorpusID:160025533}
\showURL{%
\tempurl}


\bibitem[Rozière et~al\mbox{.}(2023)]%
        {rozière2023codellama}
\bibfield{author}{\bibinfo{person}{Baptiste Rozière}, \bibinfo{person}{Jonas Gehring}, \bibinfo{person}{Fabian Gloeckle}, \bibinfo{person}{Sten Sootla}, \bibinfo{person}{Itai Gat}, \bibinfo{person}{Xiaoqing~Ellen Tan}, \bibinfo{person}{Yossi Adi}, \bibinfo{person}{Jingyu Liu}, \bibinfo{person}{Tal Remez}, \bibinfo{person}{Jérémy Rapin}, \bibinfo{person}{Artyom Kozhevnikov}, \bibinfo{person}{Ivan Evtimov}, \bibinfo{person}{Joanna Bitton}, \bibinfo{person}{Manish Bhatt}, \bibinfo{person}{Cristian~Canton Ferrer}, \bibinfo{person}{Aaron Grattafiori}, \bibinfo{person}{Wenhan Xiong}, \bibinfo{person}{Alexandre Défossez}, \bibinfo{person}{Jade Copet}, \bibinfo{person}{Faisal Azhar}, \bibinfo{person}{Hugo Touvron}, \bibinfo{person}{Louis Martin}, \bibinfo{person}{Nicolas Usunier}, \bibinfo{person}{Thomas Scialom}, {and} \bibinfo{person}{Gabriel Synnaeve}.} \bibinfo{year}{2023}\natexlab{}.
\newblock \bibinfo{title}{Code Llama: Open Foundation Models for Code}.
\newblock
\newblock
\showeprint[arxiv]{2308.12950}~[cs.CL]


\bibitem[Sanh et~al\mbox{.}(2022)]%
        {sanh2022multitask}
\bibfield{author}{\bibinfo{person}{Victor Sanh}, \bibinfo{person}{Albert Webson}, \bibinfo{person}{Colin Raffel}, \bibinfo{person}{Stephen~H. Bach}, \bibinfo{person}{Lintang Sutawika}, \bibinfo{person}{Zaid Alyafeai}, \bibinfo{person}{Antoine Chaffin}, \bibinfo{person}{Arnaud Stiegler}, \bibinfo{person}{Teven~Le Scao}, \bibinfo{person}{Arun Raja}, \bibinfo{person}{Manan Dey}, \bibinfo{person}{M~Saiful Bari}, \bibinfo{person}{Canwen Xu}, \bibinfo{person}{Urmish Thakker}, \bibinfo{person}{Shanya~Sharma Sharma}, \bibinfo{person}{Eliza Szczechla}, \bibinfo{person}{Taewoon Kim}, \bibinfo{person}{Gunjan Chhablani}, \bibinfo{person}{Nihal Nayak}, \bibinfo{person}{Debajyoti Datta}, \bibinfo{person}{Jonathan Chang}, \bibinfo{person}{Mike Tian-Jian Jiang}, \bibinfo{person}{Han Wang}, \bibinfo{person}{Matteo Manica}, \bibinfo{person}{Sheng Shen}, \bibinfo{person}{Zheng~Xin Yong}, \bibinfo{person}{Harshit Pandey}, \bibinfo{person}{Rachel Bawden}, \bibinfo{person}{Thomas Wang}, \bibinfo{person}{Trishala Neeraj},
  \bibinfo{person}{Jos Rozen}, \bibinfo{person}{Abheesht Sharma}, \bibinfo{person}{Andrea Santilli}, \bibinfo{person}{Thibault Fevry}, \bibinfo{person}{Jason~Alan Fries}, \bibinfo{person}{Ryan Teehan}, \bibinfo{person}{Tali Bers}, \bibinfo{person}{Stella Biderman}, \bibinfo{person}{Leo Gao}, \bibinfo{person}{Thomas Wolf}, {and} \bibinfo{person}{Alexander~M. Rush}.} \bibinfo{year}{2022}\natexlab{}.
\newblock \bibinfo{title}{Multitask Prompted Training Enables Zero-Shot Task Generalization}.
\newblock
\newblock
\showeprint[arxiv]{2110.08207}~[cs.LG]


\bibitem[Shinn et~al\mbox{.}(2023)]%
        {shinn2023reflexion}
\bibfield{author}{\bibinfo{person}{Noah Shinn}, \bibinfo{person}{Federico Cassano}, \bibinfo{person}{Edward Berman}, \bibinfo{person}{Ashwin Gopinath}, \bibinfo{person}{Karthik Narasimhan}, {and} \bibinfo{person}{Shunyu Yao}.} \bibinfo{year}{2023}\natexlab{}.
\newblock \bibinfo{title}{Reflexion: Language Agents with Verbal Reinforcement Learning}.
\newblock
\newblock
\showeprint[arxiv]{2303.11366}~[cs.AI]


\bibitem[Srivastava et~al\mbox{.}(2014)]%
        {srivastava2014dropout}
\bibfield{author}{\bibinfo{person}{Nitish Srivastava}, \bibinfo{person}{Geoffrey Hinton}, \bibinfo{person}{Alex Krizhevsky}, \bibinfo{person}{Ilya Sutskever}, {and} \bibinfo{person}{Ruslan Salakhutdinov}.} \bibinfo{year}{2014}\natexlab{}.
\newblock \showarticletitle{Dropout: A Simple Way to Prevent Neural Networks from Overfitting}.
\newblock \bibinfo{journal}{\emph{Journal of Machine Learning Research}} \bibinfo{volume}{15}, \bibinfo{number}{56} (\bibinfo{year}{2014}), \bibinfo{pages}{1929--1958}.
\newblock
\urldef\tempurl%
\url{http://jmlr.org/papers/v15/srivastava14a.html}
\showURL{%
\tempurl}


\bibitem[Stiennon et~al\mbox{.}(2022)]%
        {stiennon2022learning}
\bibfield{author}{\bibinfo{person}{Nisan Stiennon}, \bibinfo{person}{Long Ouyang}, \bibinfo{person}{Jeff Wu}, \bibinfo{person}{Daniel~M. Ziegler}, \bibinfo{person}{Ryan Lowe}, \bibinfo{person}{Chelsea Voss}, \bibinfo{person}{Alec Radford}, \bibinfo{person}{Dario Amodei}, {and} \bibinfo{person}{Paul Christiano}.} \bibinfo{year}{2022}\natexlab{}.
\newblock \bibinfo{title}{Learning to summarize from human feedback}.
\newblock
\newblock
\showeprint[arxiv]{2009.01325}~[cs.CL]


\bibitem[Tillmann and Ney(2003)]%
        {tillmann-ney-2003-beam-search}
\bibfield{author}{\bibinfo{person}{Christoph Tillmann} {and} \bibinfo{person}{Hermann Ney}.} \bibinfo{year}{2003}\natexlab{}.
\newblock \showarticletitle{Word Reordering and a Dynamic Programming Beam Search Algorithm for Statistical Machine Translation}.
\newblock \bibinfo{journal}{\emph{Computational Linguistics}} \bibinfo{volume}{29}, \bibinfo{number}{1} (\bibinfo{year}{2003}), \bibinfo{pages}{97--133}.
\newblock
\urldef\tempurl%
\url{https://doi.org/10.1162/089120103321337458}
\showDOI{\tempurl}


\bibitem[Touvron et~al\mbox{.}(2023)]%
        {touvron2023llama}
\bibfield{author}{\bibinfo{person}{Hugo Touvron}, \bibinfo{person}{Louis Martin}, \bibinfo{person}{Kevin Stone}, \bibinfo{person}{Peter Albert}, \bibinfo{person}{Amjad Almahairi}, \bibinfo{person}{Yasmine Babaei}, \bibinfo{person}{Nikolay Bashlykov}, \bibinfo{person}{Soumya Batra}, \bibinfo{person}{Prajjwal Bhargava}, \bibinfo{person}{Shruti Bhosale}, \bibinfo{person}{Dan Bikel}, \bibinfo{person}{Lukas Blecher}, \bibinfo{person}{Cristian~Canton Ferrer}, \bibinfo{person}{Moya Chen}, \bibinfo{person}{Guillem Cucurull}, \bibinfo{person}{David Esiobu}, \bibinfo{person}{Jude Fernandes}, \bibinfo{person}{Jeremy Fu}, \bibinfo{person}{Wenyin Fu}, \bibinfo{person}{Brian Fuller}, \bibinfo{person}{Cynthia Gao}, \bibinfo{person}{Vedanuj Goswami}, \bibinfo{person}{Naman Goyal}, \bibinfo{person}{Anthony Hartshorn}, \bibinfo{person}{Saghar Hosseini}, \bibinfo{person}{Rui Hou}, \bibinfo{person}{Hakan Inan}, \bibinfo{person}{Marcin Kardas}, \bibinfo{person}{Viktor Kerkez}, \bibinfo{person}{Madian Khabsa},
  \bibinfo{person}{Isabel Kloumann}, \bibinfo{person}{Artem Korenev}, \bibinfo{person}{Punit~Singh Koura}, \bibinfo{person}{Marie-Anne Lachaux}, \bibinfo{person}{Thibaut Lavril}, \bibinfo{person}{Jenya Lee}, \bibinfo{person}{Diana Liskovich}, \bibinfo{person}{Yinghai Lu}, \bibinfo{person}{Yuning Mao}, \bibinfo{person}{Xavier Martinet}, \bibinfo{person}{Todor Mihaylov}, \bibinfo{person}{Pushkar Mishra}, \bibinfo{person}{Igor Molybog}, \bibinfo{person}{Yixin Nie}, \bibinfo{person}{Andrew Poulton}, \bibinfo{person}{Jeremy Reizenstein}, \bibinfo{person}{Rashi Rungta}, \bibinfo{person}{Kalyan Saladi}, \bibinfo{person}{Alan Schelten}, \bibinfo{person}{Ruan Silva}, \bibinfo{person}{Eric~Michael Smith}, \bibinfo{person}{Ranjan Subramanian}, \bibinfo{person}{Xiaoqing~Ellen Tan}, \bibinfo{person}{Binh Tang}, \bibinfo{person}{Ross Taylor}, \bibinfo{person}{Adina Williams}, \bibinfo{person}{Jian~Xiang Kuan}, \bibinfo{person}{Puxin Xu}, \bibinfo{person}{Zheng Yan}, \bibinfo{person}{Iliyan Zarov}, \bibinfo{person}{Yuchen
  Zhang}, \bibinfo{person}{Angela Fan}, \bibinfo{person}{Melanie Kambadur}, \bibinfo{person}{Sharan Narang}, \bibinfo{person}{Aurelien Rodriguez}, \bibinfo{person}{Robert Stojnic}, \bibinfo{person}{Sergey Edunov}, {and} \bibinfo{person}{Thomas Scialom}.} \bibinfo{year}{2023}\natexlab{}.
\newblock \bibinfo{title}{Llama 2: Open Foundation and Fine-Tuned Chat Models}.
\newblock
\newblock
\showeprint[arxiv]{2307.09288}~[cs.CL]


\bibitem[Vaswani et~al\mbox{.}(2017)]%
        {vaswani2017transformer}
\bibfield{author}{\bibinfo{person}{Ashish Vaswani}, \bibinfo{person}{Noam Shazeer}, \bibinfo{person}{Niki Parmar}, \bibinfo{person}{Jakob Uszkoreit}, \bibinfo{person}{Llion Jones}, \bibinfo{person}{Aidan~N. Gomez}, \bibinfo{person}{Lukasz Kaiser}, {and} \bibinfo{person}{Illia Polosukhin}.} \bibinfo{year}{2017}\natexlab{}.
\newblock \showarticletitle{Attention is All You Need}. In \bibinfo{booktitle}{\emph{Proceedings of the 31st International Conference on Neural Information Processing Systems}} (Long Beach, California, USA) \emph{(\bibinfo{series}{NIPS'17})}. \bibinfo{pages}{6000–6010}.
\newblock
\showISBNx{9781510860964}


\bibitem[Wang et~al\mbox{.}(2021)]%
        {wang2021selftuning}
\bibfield{author}{\bibinfo{person}{Ximei Wang}, \bibinfo{person}{Jinghan Gao}, \bibinfo{person}{Mingsheng Long}, {and} \bibinfo{person}{Jianmin Wang}.} \bibinfo{year}{2021}\natexlab{}.
\newblock \showarticletitle{Self-Tuning for Data-Efficient Deep Learning}. In \bibinfo{booktitle}{\emph{International Conference on Machine Learning (ICML)}}.
\newblock


\bibitem[Wang and Schneider(2015)]%
        {wang2015generalization}
\bibfield{author}{\bibinfo{person}{Xuezhi Wang} {and} \bibinfo{person}{Jeff Schneider}.} \bibinfo{year}{2015}\natexlab{}.
\newblock \showarticletitle{Generalization Bounds for Transfer Learning under Model Shift}. In \bibinfo{booktitle}{\emph{Proceedings of the Thirty-First Conference on Uncertainty in Artificial Intelligence}} (Amsterdam, Netherlands) \emph{(\bibinfo{series}{UAI'15})}. \bibinfo{publisher}{AUAI Press}, \bibinfo{address}{Arlington, Virginia, USA}, \bibinfo{pages}{922–931}.
\newblock
\showISBNx{9780996643108}


\bibitem[Wei et~al\mbox{.}(2022a)]%
        {wei2022finetuned}
\bibfield{author}{\bibinfo{person}{Jason Wei}, \bibinfo{person}{Maarten Bosma}, \bibinfo{person}{Vincent~Y. Zhao}, \bibinfo{person}{Kelvin Guu}, \bibinfo{person}{Adams~Wei Yu}, \bibinfo{person}{Brian Lester}, \bibinfo{person}{Nan Du}, \bibinfo{person}{Andrew~M. Dai}, {and} \bibinfo{person}{Quoc~V. Le}.} \bibinfo{year}{2022}\natexlab{a}.
\newblock \bibinfo{title}{Finetuned Language Models Are Zero-Shot Learners}.
\newblock
\newblock
\showeprint[arxiv]{2109.01652}~[cs.CL]


\bibitem[Wei et~al\mbox{.}(2022b)]%
        {wei2022emergent}
\bibfield{author}{\bibinfo{person}{Jason Wei}, \bibinfo{person}{Yi Tay}, \bibinfo{person}{Rishi Bommasani}, \bibinfo{person}{Colin Raffel}, \bibinfo{person}{Barret Zoph}, \bibinfo{person}{Sebastian Borgeaud}, \bibinfo{person}{Dani Yogatama}, \bibinfo{person}{Maarten Bosma}, \bibinfo{person}{Denny Zhou}, \bibinfo{person}{Donald Metzler}, \bibinfo{person}{Ed~H. Chi}, \bibinfo{person}{Tatsunori Hashimoto}, \bibinfo{person}{Oriol Vinyals}, \bibinfo{person}{Percy Liang}, \bibinfo{person}{Jeff Dean}, {and} \bibinfo{person}{William Fedus}.} \bibinfo{year}{2022}\natexlab{b}.
\newblock \bibinfo{title}{Emergent Abilities of Large Language Models}.
\newblock
\newblock
\showeprint[arxiv]{2206.07682}~[cs.CL]


\bibitem[West et~al\mbox{.}(2022)]%
        {west-etal-2022-symbolic}
\bibfield{author}{\bibinfo{person}{Peter West}, \bibinfo{person}{Chandra Bhagavatula}, \bibinfo{person}{Jack Hessel}, \bibinfo{person}{Jena Hwang}, \bibinfo{person}{Liwei Jiang}, \bibinfo{person}{Ronan Le~Bras}, \bibinfo{person}{Ximing Lu}, \bibinfo{person}{Sean Welleck}, {and} \bibinfo{person}{Yejin Choi}.} \bibinfo{year}{2022}\natexlab{}.
\newblock \showarticletitle{Symbolic Knowledge Distillation: from General Language Models to Commonsense Models}. In \bibinfo{booktitle}{\emph{Proceedings of the 2022 Conference of the North American Chapter of the Association for Computational Linguistics: Human Language Technologies}}. \bibinfo{publisher}{Association for Computational Linguistics}, \bibinfo{address}{Seattle, United States}, \bibinfo{pages}{4602--4625}.
\newblock
\urldef\tempurl%
\url{https://doi.org/10.18653/v1/2022.naacl-main.341}
\showDOI{\tempurl}


\bibitem[Xu et~al\mbox{.}(2022)]%
        {xu2022polycoder}
\bibfield{author}{\bibinfo{person}{Frank~F. Xu}, \bibinfo{person}{Uri Alon}, \bibinfo{person}{Graham Neubig}, {and} \bibinfo{person}{Vincent~Josua Hellendoorn}.} \bibinfo{year}{2022}\natexlab{}.
\newblock \showarticletitle{A Systematic Evaluation of Large Language Models of Code}. In \bibinfo{booktitle}{\emph{Proceedings of the 6th ACM SIGPLAN International Symposium on Machine Programming}} (San Diego, CA, USA) \emph{(\bibinfo{series}{MAPS 2022})}. \bibinfo{publisher}{Association for Computing Machinery}, \bibinfo{address}{New York, NY, USA}, \bibinfo{pages}{1–10}.
\newblock
\showISBNx{9781450392730}
\urldef\tempurl%
\url{https://doi.org/10.1145/3520312.3534862}
\showDOI{\tempurl}


\bibitem[Zan et~al\mbox{.}(2023)]%
        {zan-etal-2023-large}
\bibfield{author}{\bibinfo{person}{Daoguang Zan}, \bibinfo{person}{Bei Chen}, \bibinfo{person}{Fengji Zhang}, \bibinfo{person}{Dianjie Lu}, \bibinfo{person}{Bingchao Wu}, \bibinfo{person}{Bei Guan}, \bibinfo{person}{Wang Yongji}, {and} \bibinfo{person}{Jian-Guang Lou}.} \bibinfo{year}{2023}\natexlab{}.
\newblock \showarticletitle{Large Language Models Meet {NL}2{C}ode: A Survey}. In \bibinfo{booktitle}{\emph{Proceedings of the 61st Annual Meeting of the Association for Computational Linguistics (Volume 1: Long Papers)}}. \bibinfo{publisher}{Association for Computational Linguistics}, \bibinfo{address}{Toronto, Canada}, \bibinfo{pages}{7443--7464}.
\newblock
\urldef\tempurl%
\url{https://doi.org/10.18653/v1/2023.acl-long.411}
\showDOI{\tempurl}


\bibitem[Zhang et~al\mbox{.}(2023)]%
        {zhang2023selfedit}
\bibfield{author}{\bibinfo{person}{Kechi Zhang}, \bibinfo{person}{Zhuo Li}, \bibinfo{person}{Jia Li}, \bibinfo{person}{Ge Li}, {and} \bibinfo{person}{Zhi Jin}.} \bibinfo{year}{2023}\natexlab{}.
\newblock \bibinfo{title}{Self-Edit: Fault-Aware Code Editor for Code Generation}.
\newblock
\newblock
\showeprint[arxiv]{2305.04087}~[cs.SE]


\bibitem[Ziegler et~al\mbox{.}(2020)]%
        {ziegler2020finetuning}
\bibfield{author}{\bibinfo{person}{Daniel~M. Ziegler}, \bibinfo{person}{Nisan Stiennon}, \bibinfo{person}{Jeffrey Wu}, \bibinfo{person}{Tom~B. Brown}, \bibinfo{person}{Alec Radford}, \bibinfo{person}{Dario Amodei}, \bibinfo{person}{Paul Christiano}, {and} \bibinfo{person}{Geoffrey Irving}.} \bibinfo{year}{2020}\natexlab{}.
\newblock \bibinfo{title}{Fine-Tuning Language Models from Human Preferences}.
\newblock
\newblock
\showeprint[arxiv]{1909.08593}~[cs.CL]


\end{thebibliography}

\end{document}